\documentclass[%
 aip,
amsmath,amssymb,
preprint,%
]{revtex4-2}

\usepackage{graphicx}
\usepackage{dcolumn}
\usepackage{bm}
\usepackage{xcolor}
\usepackage[utf8]{inputenc}
\usepackage[T1]{fontenc}
\usepackage{mathptmx}
\usepackage{etoolbox}
\usepackage[nice]{nicefrac}
\makeatletter
\def\@email#1#2{%
 \endgroup
 \patchcmd{\titleblock@produce}
  {\frontmatter@RRAPformat}
  {\frontmatter@RRAPformat{\produce@RRAP{*#1\href{mailto:#2}{#2}}}\frontmatter@RRAPformat}
  {}{}
}%

\makeatother
\begin{document}

\preprint{AIP/123-QED}
\title{Assessing transition rates as functions of environmental variables}
\author{Luca Donati}
\email{donati@zib.de}
\affiliation{Zuse Institute Berlin, Takustr. 7, D-14195 Berlin, Germany}
\author{Marcus Weber}
\email{weber@zib.de}
\affiliation{Zuse Institute Berlin, Takustr. 7, D-14195 Berlin, Germany}

\date{\today}

\begin{abstract}
We present a method to estimate the transition rates of molecular systems under different environmental conditions which cause the formation or the breaking of bonds and require the sampling of the Grand Canonical Ensemble.
For this purpose, we model the molecular system in terms of probable ``scenarios'', governed by different potential energy functions, which are separately sampled by classical MD simulations.
Reweighting the canonical distribution of each scenario according to specific environmental variables, we estimate the grand canonical distribution, then we use the Square Root Approximation (SqRA) method to discretize the Fokker-Planck operator into a rate matrix and the robust Perron Cluster Cluster Analysis (PCCA+) method to coarse-grain the kinetic model.
This permits to efficiently estimate the transition rates of conformational states as functions of environmental variables, for example, the local pH at a cell membrane.
In this work we formalize the theoretical framework of the procedure and we present a numerical experiment comparing the results with those provided by a constant-pH method based on non-equilibrium Molecular Dynamics Monte Carlo simulations.
The method is relevant for the development of new drug design strategies which take into account how the cellular environment influences biochemical processes.
\end{abstract}

\maketitle
%
%
\section{Introduction} 
The comprehension at the microscopic level of the biological processes taking place within the cell is fundamental to understand the origin of diseases and to develop more accurate and effective pharmaceutical products.
This is pursued also by means of Molecular Dynamics (MD) simulations \cite{Haile1997, Frenkel2002}, where the conformational ensemble and the dynamics of biomolecules are sampled integrating the equations which govern the motion of the atoms. 
However, the environment of the cell plays a determinant role for the structure and the kinetics properties of biomolecules and, thus, for their function.
For example, the acidity of the cell influences the dynamics of ligand-receptor systems and must be taken into account in a drug design protocol \cite{Manallack2013}.

Unfortunately, simulating the environmental impact on large  biomolecules is not feasible as quantum mechanical mechanisms are involved, for example the bond breaking and bond formation due to a change in pH.
These mechanisms are not allowed in a standard classical MD simulation, thus, over the last two decades, several methods and strategies, based on different theoretical hypothesis, have been proposed to simulate the dynamics of molecules at constant-pH, where titratable sites of biomolecules can be protonated or deprotonated during the simulation \cite{Mongan2005, Chen2014, Barroso2017}.

A common strategy to perform constant-pH simulations is to combine classical MD with Monte Carlo (MC), where the molecule is allowed or not allowed to change its protonation state according to a Metropolis criterion.
One of the first hybrid MDMC approach was proposed by B\"urgi et al. \cite{Burgi2002} where a thermodynamic integration was used to estimate the free energy differences between states.
To improve the accuracy and to reduce the computational effort, several solutions and improvements have been later suggested  \cite{Mongan2004,Williams2010,Baptista2002,Meng2010,Dashti2012,Swails2014}.
Recently it was proposed by Y.~Chen et al.~\cite{Chen2015} to use a non-equilibrium Molecular Dynamics Monte Carlo (neMDMC) method, where the transition between protonated/unprotonated states is performed by short non equilibrium simulations, accepted or rejected according to a Metropolis algorithm.
The neMDMC approach, based on the work of H.~Stern \cite{Stern2007}, was further developed to study large biomolecules with several titratable sites \cite{Radak2017}.

Typically, the hybrid MDMC approach is used to estimate the titration curves, but they could be also employed to estimate transition rates.
This requires building  a Markov State Model (MSM) \cite{Schuette1999b,  Deuflhard2000, Chodera2007, Buchete2008, Prinz2011, Keller2011, keller2019}, i.e. a representation of the dynamics in terms of jump probabilities between subsets of the state space which approximates the underlying transfer operator.
%
%
Afterward, applying coarse-graining techniques such as the robust Perron Cluster Cluster Analysis (PCCA+) \cite{Deuflhard2004, Kube2007, Roblitz2013, Weber2018, Erlekam2019}, one identifies the metastabilities and builds a coarse-grained transition probability matrix which, in the case of a ligand-receptor system, represents the dynamics between the bound and unbound macrostate.

From a statistical thermodynamic point of view, MDMC simulations sample the Grand Canonical Ensemble (GCE), i.e. the collection of microstates of the system where the number of atoms varies in order to keep the chemical potential of the components constant.
The GCE is the sum of Canonical Ensembles (CEs) opportunely weighted by an exponential factor which depends by the chemical potential under specific environmental conditions. 

On the basis of this insight, here we propose a different strategy that we refer to as ``GCEkinCEs'' as it allows for extracting Kinetics information of the GCE from the distributions of the CEs.
Instead of sampling directly the GCE through direct non-equilibrium simulations, we sample separately the canonical ensembles via classical MD simulations, then we reweight and sum the canonical distributions to obtain the grand canonical distribution as function of the environmental variables under investigation, for example the environmental pH.
One of the limits of this approach is the high number of possible canonical ensembles that could emerge changing the atom numbers of each component.
As an alternative, we redefine the grand canonical ensemble in terms of possible \emph{scenarios}, i.e. we consider only those configurations with a high probability to occur.
After having estimated the grand canonical distribution, we employ the Square Root Approximation (SqRA) of the Fokker-Planck operator, a method that approximates the rate between two adjacent subsets of the state space as a geometric average of their Boltzmann weights multiplied by a flux term.
The SqRA formula was obtained in several contexts.
First, Bicout and Szabo, applying a finite different scheme, derived the formula for a one-dimensional system to study the dynamics of electron transfer in a non-Debye solvent \cite{Bicout1998}.
An analogous formula for high-dimensional systems was derived from the continuity equation exploiting Gauss's flux theorem \cite{Lie2013, Donati2018b}.
A third derivation exploits the maximum caliber (maximum path entropy) approach \cite{Dixit2015, Stock2008, Otten2010}.
In addition, SqRA was implemented by Rosta and Hummer \cite{Rosta2015} into the Dynamic Histogram Analysis Method (DHAM) for estimate transition probabilities with small lag time from umbrella sampling simulations.
Recently the method has been further expanded by determining the correct expression of the flux term for high-dimensional systems \cite{Donati2021}.
SqRA yields a fine-grained rate matrix whose eigenvectors are used to generate a coarse-grained model via PCCA+ which provides the transition rates between macrostates.

The proposed strategy was already used by S.~Ray et al.~\cite{Ray2020} to estimate the binding rates as functions of the environmental pH of the opioids fentanyl and N-(3-fluoro-1-phenethylpiperidin-4-yl)-N-phenyl propionamide (NFEPP) \cite{Spahn2017} in a $\mu$-opioid receptor.
Here, we formalize the procedure providing the theoretical framework and we generalize the method to the case of $k$ environmental variables which influence the dynamics of the system, for example the salt or ion concentration.
Furthermore, we compare GCEkinCEs with the constant-pH MD simulation method proposed by Y.~Chen \cite{Chen2015}, showing that the methods are able to reproduce the same grand canonical distribution.

The paper is outlined as follows:
\begin{enumerate}
    \item In section \ref{sec:theory} we recall the theory of the grand canonical ensemble, we explain how to reduce the dimensionality of the grand canonical partition function and the grand canonical distribution.
    Then, we introduce the equations of motion, discuss the dimensionality reduction and the Fokker-Planck equation in the reduced space.
    \item In section \ref{sec:methods} we explain the procedure of GCEkinCEs and the techniques  involved, focusing on the SqRA and the PCCA+ methods.
    \item In section \ref{sec:results} we present a numerical experiment and we compare the results with those obtained from neMDMC simulations. 
    \item In section \ref{sec:conclusion} we present our conclusions and the outlook.
\end{enumerate}
%
\begin{figure}[ht]
    \centering
    \includegraphics[scale=1]{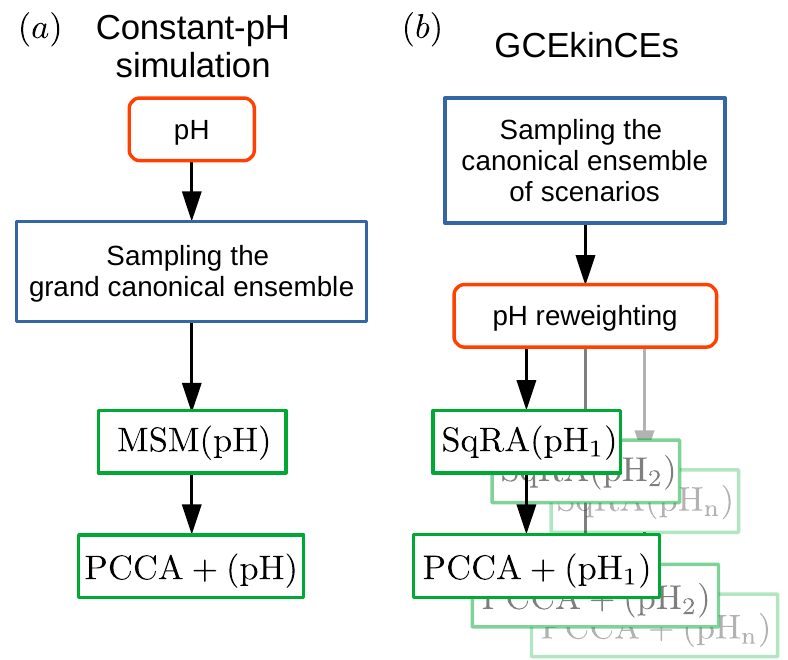}
    \caption
    {
    Workflow diagram of the methods: (a)  Constant-pH simulations directly sample the GCE; (b) GCEkinCEs samples the CEs of probable scenarios and reweights the grand canonical distribution for several pH values.
    } 
    \label{fig:fig0}
\end{figure}
%
\newpage
\section{Theoretical background}
\label{sec:theory}
\subsection{The Grand Canonical Ensemble}
Consider a system in thermal equilibrium with a reservoir which maintains a constant temperature $T$ and a constant number of number of atoms $N$ in a fixed volume $V$.
The Hamiltonian of the system is given by
\begin{eqnarray}
\mathcal{H}(\mathbf{r}, \mathbf{p}) 
&=&
\frac{1}{2} \mathbf{p}^\top \mathbf{M}^{-1} \mathbf{p}
+
U(\mathbf{r}) \, ,
\end{eqnarray}
where $\mathbf{M}$ is a $3N\times 3N$ diagonal matrix containing the mass of the atoms and $U(\mathbf{r})$ is the potential energy function.
The vectors $\mathbf{r}$ and $\mathbf{p}$ denote respectively the Cartesian coordinates and the momenta of the atoms, and together constitute the state of the system $\mathbf{x} = \lbrace\mathbf{r}, \mathbf{p}\rbrace \in \Gamma_{\mathbf{r}} \times \Gamma_{\mathbf{p}} = \Gamma \subset \mathbb{R}^{6n}$, where $\Gamma$ is the phase space.

The collection of all possible states so delineated  is called Canonical Ensemble (CE) and the corresponding partition function is defined as
\begin{eqnarray}
Z(N,V,T) =  \iint_{\mathbb{R}^{6N}} \mathrm{d}\mathbf{r} \mathrm{d}\mathbf{p}\, e^{-\beta \mathcal{H}(\mathbf{r}, \mathbf{p})}\, ,
\end{eqnarray}
where $\beta = \nicefrac{1}{k_B T}$ with $k_B$ the Boltzmann constant.

Consider now the case where the number of atoms is not fixed, but varies due to a change of the environmental conditions, then the CE is generalized by the Grand Canonical Ensemble (GCE), whose partition function is written as
\begin{eqnarray}
\mathcal{Z}(\mu,V,T) 
&=& 
\sum_{N=0}^{\infty}
\iint_{\mathbb{R}^{6N}} \mathrm{d}\mathbf{r} \mathrm{d}\mathbf{p} \, 
e^{\beta \left( 
N\mu - \mathcal{H}_N(\mathbf{r},\mathbf{p})
\right)}  \cr
&=& 
\sum_{N=0}^{\infty} e^{\beta N\mu} 
\iint_{\mathbb{R}^{6N}} \mathrm{d}\mathbf{r} \mathrm{d}\mathbf{p} \, e^{-\beta \mathcal{H}_N(\mathbf{r},\mathbf{p})}  \cr
&=& 
\sum_{N=0}^{\infty} e^{\beta N\mu} Z(N,V,T)
 \, ,
\label{eq:GrandCanEnsemble1}
\end{eqnarray}
where $\mathcal{H}_{N}(\mathbf{r},\mathbf{p})$ is the Hamiltonian associated to the system state with $N$ atoms, and $\mu$ is the chemical potential defined by the linear relation
\begin{eqnarray}
\mu & = & \mu_0 + N_A k_B T \log \frac{c}{c_0} \cr 
&=& \mu_0 + RT \log a \, ,  
\end{eqnarray}
where $\mu_0$ and $c_0$ are respectively the standard chemical potential and the standard concentration, and $N_A$ is the Avogadro constant.
The unit-less quantity $a$ is the concentration activity, i.e. the effective concentration of substance which reacts.
For example, the proton chemical potential, assuming $\mu_0=0$ is
\begin{eqnarray}
\mu_{\mathrm{H+}} = - R T \log(10) \, \mathrm{pH} \, ,
\end{eqnarray}
where we used the definition of $\mathrm{pH} = - \log_{10} [\mathrm{H+}]$.

The chemical potential represents the energy absorbed or released by the system when the particle number changes.
In other words, it tells which system state is more likely to occur and expresses the statistical weight of each canonical partition function $Z(N,V,T)$. 

Conversely from other environmental parameters, such as the temperature $T$, the chemical potential is not unique, but related to each component of the system.
Thus, if the system has $k$ components, the chemical potentials can be described by a vector $\Vec{\mu} = (\mu_1,...,\mu_k)$ and the GCE is generalized as
\begin{eqnarray}
\mathcal{Z}(\vec{\mu},V,T) 
&=& 
\sum_{N_1=0}^{\infty} 
\cdot \dotsc \cdot  
\sum_{N_k=0}^{\infty} 
\iint_{\mathbb{R}^{6N}} \mathrm{d}\mathbf{r} \mathrm{d}\mathbf{p} \, 
\exp\left[
\beta 
\left(\sum_{i=1}^{k}N_i\mu_i
 -\mathcal{H}_{\vec{N}}(\mathbf{r},\mathbf{p})
 \right)
 \right] \cr
&=& 
\sum_{N_1=0}^{\infty} 
\cdot \dotsc \cdot  
\sum_{N_k=0}^{\infty} 
\exp\left(
\beta \sum_{i=1}^{k}N_i\mu_i
\right)
\iint_{\mathbb{R}^{6N}} \mathrm{d}\mathbf{r} \mathrm{d}\mathbf{p} \, 
e^{ -\beta \mathcal{H}_{\vec{N}}(\mathbf{r},\mathbf{p})}
\cr
&=& 
\sum_{N_1=0}^{\infty} e^{\beta N_1\mu_1 } 
\cdot \dotsc \cdot  
\sum_{N_k=0}^{\infty} 
e^{\beta N_k\mu_k }
\iint_{\mathbb{R}^{6N}} \mathrm{d}\mathbf{r} \mathrm{d}\mathbf{p} \, 
e^{ -\beta \mathcal{H}_{\vec{N}}(\mathbf{r},\mathbf{p})}
\cr
&=& 
\prod_{i=1}^{k} 
\sum_{N_i=0}^{\infty} 
e^{\beta N_i \mu_i}
\iint_{\mathbb{R}^{6N}} \mathrm{d}\mathbf{r} \mathrm{d}\mathbf{p} \, 
e^{ -\beta \mathcal{H}_{\vec{N}}(\mathbf{r},\mathbf{p})}
\cr
&=& 
\prod_{i=1}^{k} 
\sum_{N_i=0}^{\infty} 
\iint_{\mathbb{R}^{6N}} \mathrm{d}\mathbf{r} \mathrm{d}\mathbf{p} \, 
e^{\beta \left( 
N_i\mu_i - \mathcal{H}_{\vec{N}}(\mathbf{r},\mathbf{p})
\right)}
\cr
&=& 
\prod_{i=1}^{k} \sum_{N_i=0}^{\infty} e^{\beta N_i\mu_i} Z(N_i,V,T) \cr
&=& 
\prod_{i=1}^{k}  \mathcal{Z}_i(\mu_i,V,T) \, ,
\label{eq:GrandCanPartFunc2}
\end{eqnarray} 
where we have introduced the Hamiltonian function $\mathcal{H}_{\vec{N}}(\mathbf{r},\mathbf{p})$ associated to a specific combination of atom numbers $\vec{N} = (N_1,...,N_k)$, with $N_i$ denoting the number of atoms of the $i$th component.

The grand canonical partition function so defined takes into account each microstate defined by every possible combination of atom numbers.
However, it is more convenient to take into account only configurations that are likely to be observed.
For this purpose, we define the concept of \emph{scenario}, i.e. a particular configuration of the system with a high probability to occur under a specific chemical environment situation, defined by a set of environmental variables $\vec{e}=(e_1,...,e_k)$.

Thus, given $s$ possible scenarios, the GCE defined in eq.~\ref{eq:GrandCanPartFunc2} is approximated as
\begin{eqnarray}
\mathcal{Z}(\vec{e},V,T) 
&\approx& 
\sum_{n=1}^{s}
\iint_{\mathbb{R}^{6N}} \mathrm{d}\mathbf{r} \mathrm{d}\mathbf{p} \, 
e^{\beta \left( 
\mu_n(\vec{e}) - \mathcal{H}_n(\mathbf{r},\mathbf{p})
\right)} 
\cr
&=& 
\sum_{n=1}^{s} e^{\beta \mu_n(\vec{e})} 
\iint_{\mathbb{R}^{6N}} \mathrm{d}\mathbf{r} \mathrm{d}\mathbf{p} \, e^{-\beta \mathcal{H}_n(\mathbf{r},\mathbf{p})}  \cr
&=& 
\sum_{n=1}^{s} e^{\beta \mu_n(\vec{e})} Z(n,V,T) 
\, ,
\label{eq:GrandCanPartFunc3}
\end{eqnarray}
where $Z(n,V,T)$ is the CE of the $n$th scenario and $\mathcal{H}_n(\mathbf{r},\mathbf{p})$ is the Hamiltonian function of the $n$th scenario with potential $U_n(\mathbf{r})$.
The quality of the approximation depends on the choice of scenarios considered.

In this formulation the GCE explicitly depends on the set of environmental variables $\vec{e}=(e_1,...,e_k)$ which determines which scenario is more likely to occur.
The chemical potential $\mu_n(\vec{e})$ represents now the absorbed or released energy due to a change of scenario and determines the probability
\begin{eqnarray}
w_n(\vec{e}) = e^{\beta \mu_n(\vec{e})} \, ,
\label{eq:ScenarioWeight}
\end{eqnarray}
of observing the $n$th scenario given a specific set of environmental variables.
The statistical weights must satisfy the condition 
\begin{eqnarray}
\sum_{n=1}^s w_n(\vec{e}) = 1 \,.
\label{eq:WeightsCondition}
\end{eqnarray}

Inserting eq.~\ref{eq:ScenarioWeight} into eq.~\ref{eq:GrandCanPartFunc3} yields an alternative expression for the grand canonical partition function:
\begin{eqnarray}
\mathcal{Z}(\vec{e},V,T) 
&\approx& 
\sum_{n=1}^{s}
w_n(\vec{e})
\iint_{\mathbb{R}^{6N}} \mathrm{d}\mathbf{r} \mathrm{d}\mathbf{p} \, 
e^{-\beta \mathcal{H}_n(\mathbf{r},\mathbf{p})}  \, .
\label{eq:GrandCanPartFunc4}
\end{eqnarray}
We finally obtain the grand canonical distribution 
\begin{eqnarray}
\pi(\mathbf{r}, \mathrm{p};\vec{e}) 
\approx
\frac{1}{\mathcal{Z}(\vec{e},V,T)} 
\sum_{n=1}^{s}
w_n(\vec{e})
e^{-\beta \mathcal{H}_n(\mathbf{r},\mathbf{p})} \, ,
\label{eq:GrandCanDistr}
\end{eqnarray}
which expresses the probability that a configuration $\mathbf{r}$ occurs under specific environmental conditions as weighted average of the unnormalized canonical distributions of scenarios.
%

%

\subsection{Sampling the dynamics}
\label{sec:dynamics}
Sampling the GCE with partition function defined in eq.~\ref{eq:GrandCanPartFunc4} requires defining  an extended Hamiltonian that takes into account the variation of particles number under specific environmental conditions.
This is realized, for example, by adding to the classical potential a virtual particle coupled with a heat reservoir to keep a constant temperature, and with a material reservoir, which similarly keeps constant the chemical potential \cite{Lynch1997}.
Other solutions combine MD simulations with MC steps to simulate the transition between scenarios \cite{Burgi2002,Mongan2004,Williams2010,Baptista2002,Meng2010,Dashti2012,Swails2014,Stern2007,Chen2015,Radak2017}

Our idea is to separately sample the CE of each scenario by classical MD simulation and to weight each canonical distribution according to the environmental variables of interest to get the grand canonical distribution in eq.~\ref{eq:ProbRelCoord}.
This choice is motivated by eq.~\ref{eq:GrandCanDistr} which suggests that the grand canonical distribution is composed by the canonical distributions of scenarios.
Then, we introduce the equations of motion of the $n$th scenario
\begin{eqnarray}
\dot{\mathbf{r}}_n &=&
\mathbf{M}_n^{-1}\frac{\partial \mathcal{H}_n}{\partial \mathbf{p}} \, , \cr 
\dot{\mathbf{p}}_n &=&
- \frac{\partial \mathcal{H}_n}{\partial \mathbf{r}_n} 
- \boldsymbol{\xi}_n(\mathbf{r},\mathbf{p}_n)\mathbf{p}_n
+F_{\mathrm{ext}}
\, ,
\label{eq:eqsMotion}
\end{eqnarray}
where $\boldsymbol{\xi}_n(\mathbf{r}_n,\mathbf{p}_n)$ is a $3N_n\times 3N_n$ diagonal matrix containing the friction acting on atoms due to a solvent and $F_{\mathrm{ext}}$ is an external force.
The subscript $n$ denotes the $n$th scenario.
Note that, each scenario could include a different number of atoms, so the number of equations to be integrated for each scenario could change.

The choice of $\boldsymbol{\xi}_n(\mathbf{r}_n,\mathbf{p}_n)$ and $F_{\mathrm{ext}}$ defines the thermostat that simulates the exchange of energy with a heat bath to maintain a constant temperature and determines the properties of the dynamics \cite{Schuette1999b}.
For example, if $\boldsymbol{\xi}_n(\mathbf{r}_n,\mathbf{p}_n) = 0$ and $F_{\mathrm{ext}}=0$, eqs.~\ref{eq:eqsMotion} turn to the classical Hamiltonian equations of motion which conserve the total energy, but do not sample the CE.
Instead, we require that the equations of motion generate trajectories satisfying the following properties: 
(i) Markovianity, i.e. the dynamics is described by a memoryless jump process; 
(ii) ergodicity, i.e. each state of the phase space will be visited a number of times proportional to the stationary distribution $\pi(\mathbf{r}, \mathbf{p})$ by an infinitely long trajectory; 
(iii) reversibility, i.e. it fulfills the detailed balance condition
\begin{eqnarray}
\pi(\mathbf{x}) p(\mathbf{x}, \mathbf{y}; \tau)
&=&
\pi(\mathbf{y}) p(\mathbf{y}, \mathbf{x}; \tau) \, ,
\end{eqnarray}
where we introduced the transition probability of the system $p(\mathbf{x}, \mathbf{y}; \tau)$ to go from the state $\mathbf{x}$ to the state $\mathbf{y}$ in a lag time $\tau$.
This permits us to describe the dynamics as time-evolution of a probability density $\rho(\mathbf{x}): \Gamma\rightarrow\mathbb{R}^+$ by means of a propagator, or the associated transfer operator, and to build a Markov model by utilizing the methods of SqRA and PCCA+.
For more details about the propagator formalism and the justification of these assumptions, we refer to the work of F.~N\"uske et al.~\cite{Nuske2014}.

There are several thermostats which meet these conditions \cite{Prinz2011}; on the other hand, we remark that, despite they sample the same canonical distribution, the dynamic properties extracted from the trajectories can differ from model to model \cite{Kieninger2021}.
Here, in the illustrative example, we considered the overdamped Langevin dynamics governed by the equation
\begin{eqnarray}
    \dot{\mathbf{r}}_n 
    = &-&
    \, \boldsymbol{\xi}_n^{-1} \mathbf{M}_n^{-1} \,\nabla U_n(\mathbf{r}(t)) + \sqrt{2\mathbf{D}_n} \, \dot{W}(t) \, ,
\label{eq:BrownianDynamics2}    
\end{eqnarray}
where the vector $W(t) \in \mathbb{R}^{3N_n}$ is a $3N_n$-dimensional Wiener process, and $\mathbf{D}_n=k_B T\, \boldsymbol{\xi}_n^{-1} \mathbf{M}_n^{-1}$ is a diffusion matrix. 
Note that the friction matrix $\boldsymbol{\xi}_n$ does not depend on positions and momenta in this setting.
\subsection{Effective dynamics on relevant coordinates}
The dynamics of a many-body molecular system, and its interaction with a solvent, is typically approximated by a low-dimensional model which governs the dynamics of few ``relevant coordinates'' realized as transformation of the Cartesian coordinates: $q=q(\mathbf{r}):\Gamma_{\mathbf{r}} \rightarrow \Omega \subset \mathbb{R}^d$ with $1\leq d\ll3N$, where $\Omega$ denotes the state space in the low dimensional space.
Unfortunately, the choice of relevant coordinates, and the projection of the neglected coordinates onto them, is not a trivial task.
The underlying assumption is that the complex dynamics of a molecular system gives rise to a cascade of timescales of several orders of magnitude.
This allows us to separate the slow and fast dynamics, and to formulate the dynamics of slow variables as a stochastic process, even if the original trajectory was generated by deterministic dynamics.
However, the elimination of fast variables introduces non-Markovian effects in terms of a memory kernel, which describes how the past impacts the present \cite{Mori1965,Zwanzig1973,Kampen1985}.
Over the years, different methods to estimate the memory kernel and to describe the dynamics of a molecular system by a Generalized Langevin Equation were developed \cite{straub1987, Horenko2007, darve2009, jung2017, Daldrop2018, Lee2019, Ayaz2020}.
Alternatively, it is a common practice to replace the memory kernel with a position-dependent diffusion term, calibrated to reproduce the slow timescales observed in experiments or full-atomistic simulations \cite{Best2011}.
Nonetheless, there is not guarantee that the condition of Markovianity is satisfied.
This approximation is valid only if the memory kernel quickly decays with respect to the timescales of interest.
For the purposes of this work, we use the second strategy, as we need a Markovian description of the dynamics in order to estimate the kinetic rates via SqRA and PCCA+.

We now introduce the grand canonical probability density function $\hat{\rho}(q,t;\vec{e}):\Omega \rightarrow \mathbb{R}^+$ of the relevant coordinates $q$, for specific environmental variables $\vec{e}$, whose time-evolution is governed by the Fokker-Planck equation
\begin{eqnarray}
\partial_t 
\hat{\rho}(q,t;\vec{e}) 
&=& 
\beta \nabla \left[
\hat{D}(q;\vec{e})
\nabla {\hat{A}}(q;\vec{e}) 
\,
\hat{\rho}(q,t;\vec{e}) 
\right] 
+
\Delta \left[
\hat{D}(q;\vec{e}) \, \hat{\rho}(q,t;\vec{e})\right] 
\, .
\label{eq:FP2}
\end{eqnarray}
In eq.~\ref{eq:FP2}, the symbol $\Delta = \nabla \cdot \nabla$ denotes the Laplacian of a function $f: \mathbb{R}^{3N} \rightarrow \mathbb{R}$ and the hat symbol is used to indicate any function $\hat{f}:\Omega \rightarrow \mathbb{R}$ acting on relevant coordinates.
The terms $\hat{A}(q;\vec{e}), \, \hat{D}(q;\vec{e}):\Omega \rightarrow \mathbb{R}$ denote respectively the free energy and the reduced diffusion profile, which are functions of the relevant coordinates and the environmental variables.
Eq.~\ref{eq:FP2} can also be written as
\begin{eqnarray}
\partial_t \hat{\rho}(q,t;\vec{e}) 
&=& 
\hat{\mathcal{Q}}(\vec{e}) \hat{\rho}(q;\vec{e}) \, ,
\label{eq:FPoperator}
\end{eqnarray}
where we introduced the Fokker-Planck operator $\hat{\mathcal{Q}}(\vec{e})$, i.e. the infinitesimal generator of the propagator
\begin{eqnarray}
    \hat{\mathcal{P}}(\tau;\vec{e}) = \exp \left(\hat{\mathcal{Q}}(\vec{e})\tau \right)  \, ,
\end{eqnarray}
which propagates the probability density $\hat{\rho}(q,t;\vec{e})$ forward in time by a time interval $\tau$:
\begin{eqnarray}
    \hat{\mathcal{P}}(\tau;\vec{e})\hat{\rho}(q,t;\vec{e}) = \hat{\rho}(q,t+\tau;\vec{e}) \, .
\end{eqnarray}
We remark that the operators $\hat{\mathcal{P}}(\tau;\vec{e})$ and $\hat{\mathcal{Q}}(\vec{e})$ are them self functions of the environmental variables and describe the dynamics of the grand canonical ensemble in the reduced space $\Omega$. 

Hence, in the limit of infinite time $t$, the probability density $\hat{\rho}(q,t;\vec{e})$ relaxes to the projection of the grand canonical distribution onto the relevant coordinates.
This is obtained by integrating eq.~\ref{eq:GrandCanPartFunc4} over all degrees of freedom but $q$, and normalizing by the grand canonical partition function:
\begin{eqnarray}
\hat{\pi}(q;\vec{e})  
&=& 
\frac{
\hat{\mathcal{Z}}(q;\vec{e},V,T)}
{\mathcal{Z}(\vec{e},V,T)} \cr
&=&
\frac{1}{
\mathcal{Z}(\vec{e},V,T) 
}
\sum_{n=1}^{s}
w_n(\vec{e})
\int_{\mathbb{R}^{3N}} \mathrm{d}\mathbf{r} \, 
e^{-\beta U_n(\mathbf{r})} \,\delta \left(q - q(\mathbf{r})\right)\, ,
\label{eq:ProbRelCoord}
\end{eqnarray}
where $\delta$ is the Dirac delta function.
From eq.~\ref{eq:ProbRelCoord}, one can also obtain the free energy surface on relevant coordinates $\hat{A}(q;\vec{e}):\Omega \rightarrow \mathbb{R}$:
\begin{eqnarray}
\hat{A}(q;\vec{e}) = - \frac{1}{\beta} \log \hat{\pi}(q;\vec{e})  \, .
\label{eq:FreeEnergyProfile}
\end{eqnarray}
%

%
\newpage
\section{Methods}
\label{sec:methods}
The procedure to estimate transition rates  under specific environmental conditions from the sampling of CEs, is summarized as follows:
\begin{enumerate}
    \item Generate a set of trajectories, at least one for each scenario, solving eq.~\ref{eq:eqsMotion} (sec.~\ref{subsec:EM}).
    \item Project the trajectories on a set of relevant coordinates, then compute the probability distribution of each scenario and combine them to get an approximation of the grand probability distribution of the reaction coordinates  (eq.~\ref{eq:ProbRelCoord}, sec.~\ref{subsec:voronoi}). 
    \item Discretize the Fokker-Planck operator defined into a rate matrix $\hat{\mathbf{Q}}(\vec{e})$ by SqRA (eq.~\ref{eq:FP2}, sec.~\ref{subsec:sqra}).
    \item Coarse-grain the rate matrix into a rate matrix of the conformations $\hat{\mathbf{Q}}_c(\vec{e})$ by PCCA+ (sec.~\ref{subsec:pcca}).
\end{enumerate}
In sec.~\ref{subsec:constpH}, we also present a generalization of a constant-pH MD simulation method for generic environmental variables that we use to validate our results.
%
%
\subsection{Integration scheme}
\label{subsec:EM}
The choice of the integration scheme to solve the equations of motion (eqs.~\ref{eq:eqsMotion}) depends on how the friction and the external force $F_{\mathrm{ext}}$ are defined, i.e. depends on the choice of the thermostat. %

The illustrative example discussed here is based on the overdamped Langevin dynamics (eq.~\ref{eq:BrownianDynamics2}), then we used the Euler–Maruyama integration scheme \cite{Leimkuhler2015} 
\begin{eqnarray}
    \mathbf{r}_{k+1} = \mathbf{r}_{k} - 
    \boldsymbol{\xi}_n^{-1} \mathbf{M}_n^{-1}
     \nabla U_n(\mathbf{r}_k) \Delta t + 
    \sqrt{2 \mathbf{D}_n} \sqrt{\Delta t} \, \eta_k \, ,
    \label{eq:EMscheme}
\end{eqnarray}
where $\Delta t$ is the time step, chosen smaller than the period of the fastest vibration, $\mathbf{r}_{k}$ is the vector with the atoms coordinates at time $t = k\cdot \Delta t$, and $\eta_k$ is a vector of independent and identically distributed random numbers drawn from a Gaussian distribution with zero mean and unit variance at each timestep $k$.

The Euler-Maruyama scheme is a common choice within MD community to produce time-discretized trajectories of the Brownian dynamics, but other schemes such as the Milstein algorithm  can be considered \cite{Milstein2004}.
A different dynamics in eq. \ref{eq:eqsMotion} necessitates different integrator schemes \cite{Leimkuhler2015}.
\subsection{Discretization of the state space}
\label{subsec:voronoi}
After having generated a set of trajectories, at least one for each scenario under consideration, and assuming that the dynamics of its projection onto the relevant coordinates $q$ can be described by eq.~\ref{eq:FP2}, the reduced space $\Omega$ is partitioned into $K$ disjoint subsets $\Omega_i$ such that $\Omega = \cup_{i=1}^K \Omega_i$.
Each cell $\Omega_i$ is defined by the characteristic function
\begin{eqnarray}
    \mathbf{1}_i(q) &=&
    \begin{cases}
        1   &\mathrm{if} \, q \in \Omega_i \cr
        0   &\mathrm{otherwise} \, .
    \end{cases}
\label{eq:characteristic_fct}    
\end{eqnarray}

In our numerical experiment, we used the $K$-means algorithm \cite{MacQueen1967} to find $K$ representative points that define the centers of the Voronoi cells $\Omega_i$, but other solutions could be adopted.
Note that, to obtain a partition of the space $\Omega$ that encompasses the grand canonical ensemble, we applied the $K$-means algorithm to the trajectories of the scenarios together.
\subsection{Square Root Approximation of the Fokker-Planck operator}
\label{subsec:sqra}
As a starting point for the discretization of the Fokker-Planck equation, we introduce the weighted scalar product 
\begin{eqnarray}
\langle u | v\rangle_{\hat{\pi}} 
&=& \int_{\Omega} u(q) \, v(q)\, \hat{\pi}(q;\vec{e}) \, \mathrm{d}q \, ,
\end{eqnarray}
where $u$ and $v$ are functions defined on the space $\Omega$ of the relevant coordinates, and $\hat{\pi}(q;\vec{e})$ is the grand canonical distribution  defined in eq.~\ref{eq:ProbRelCoord}.

Then, given a crisp discretization of the space, the Galerkin discretization of the Fokker-Planck operator $\hat{\mathcal{Q}}(\vec{e})$ (eq.~\ref{eq:FPoperator}) is the rate matrix $\hat{\mathbf{Q}}(\vec{e})$ with entries
\begin{eqnarray}
\hat{Q}_{ij}(\vec{e})
&=&
\langle\mathbf{1}_i|\mathbf{1}_i \rangle_{\hat{\pi}}^{-1}
\langle \mathbf{1}_j|\hat{\mathcal{Q}}(\vec{e})\mathbf{1}_i \rangle_{\hat{\pi}} \cr
&=&
\frac{1}{\hat{\pi}_i} \langle \mathbf{1}_j | \hat{\mathcal{Q}}(\vec{e}) \mathbf{1}_i \rangle_{\hat{\pi}}\,.
\label{eq:GalerkinDiscretization}
\end{eqnarray}
In eq.~\ref{eq:GalerkinDiscretization}, the entry $\hat{Q}_{ij}(\vec{e})$, with $i\not= j$, denotes the rate from cell $\Omega_i$ to cell $\Omega_j$.
As the transition rates between non-adjacent cells is zero, we have that
\begin{eqnarray}
    \hat{Q}_{ij}(\vec{e}) &=& 0 \qquad \mbox{if $i\ne j$, and $\Omega_i$ is not adjacent to $\Omega_j$} \, .
\end{eqnarray}
Conversely, for adjacent cells $\Omega_i \sim \Omega_j$,  eq.~\ref{eq:GalerkinDiscretization} takes the form \cite{Lie2013, Donati2018b}
\begin{equation}
\hat{Q}_{ij}(\vec{e}) = \frac{1}{\hat{\pi}_i} \oint_{\partial\Omega_i \partial\Omega_j} \Phi(q;\vec{e}) \,  \hat{\pi}(q;\vec{e}) \, \mathrm{d} S(q) \, , 
\label{eq:GalerkinDiscretization2}
\end{equation}
where $\oint$ denotes a surface integral over the common surface $\partial \Omega_i \partial \Omega_j$ between the cell $\Omega_i$ and $\Omega_j$.

The term $\Phi(q;\vec{e})$ is the flux through the intersecting surface, which, if assumed constant over $\partial\Omega_i \partial\Omega_j$, can be taken out of the integral and approximated as \cite{Donati2021}
\begin{eqnarray}
\Phi(q;\vec{e})
\approx
\frac{
\hat{D}_{ij}(\vec{e})
}{d_{ij}}
=
\frac{\hat{D}(q_i;\vec{e}) + \hat{D}(q_j;\vec{e})}{2d_{ij}} \, ,
\label{eq:flux}
\end{eqnarray}
where $d_{ij}$ is the Euclidean distance between the Voronoi cells centers $q_i$ and $q_j$.

The remaining part of the surface integral in eq.~\ref{eq:GalerkinDiscretization2} expresses the density on the intersecting surface $\partial \Omega_i \partial \Omega_j$:
\begin{eqnarray}
\hat{\pi}_{ij}(\vec{e})
&=&
\oint_{\partial\Omega_i \partial\Omega_j} \hat{\pi}(q;\vec{e}) \, \mathrm{d} S(q) 
\, .
\end{eqnarray}
Applying the same assumption used for the diffusion, it can be approximated as 
\begin{eqnarray}
\hat{\pi}_{ij}(\vec{e})
&\propto& 
\mathcal{S}_{ij}
\exp\left(-\beta \hat{A}(q_{ij};\vec{e}) \right) \, ,
\label{eq:piSurface}
\end{eqnarray}
where $\mathcal{S}_{ij}$ is the area of the intersecting surface, and the density is replaced by the exponential of the free energy (eq.~\ref{eq:FreeEnergyProfile}) along the intersecting surface.
In eq.~\ref{eq:piSurface}, we use the symbol $\propto$, because we omit the normalization constant.
Next, assuming that each cell is small enough such that the free energy profile $\hat{A}(q;\vec{e})$ is constant within a given cell and equal to its value $A(q_i)$ measured at the center $q_i$ of the cell, we approximate the free energy along the intersecting surface as
\begin{eqnarray}
\hat{A}(q_{ij} ;\vec{e})
&\approx& 
\frac{\hat{A}(q_i;\vec{e}) + \hat{A}(q_j;\vec{e})}{2} \cr
&\approx& 
\frac{\hat{A}(q_i;\vec{e}) + \hat{A}(q_j;\vec{e})}{2} \, .
\end{eqnarray}
Inserting this last expression into eq.~\ref{eq:piSurface} yields
\begin{eqnarray}
\hat{\pi}_{ij}(\vec{e}) 
&\propto& 
\mathcal{S}_{ij}
\exp\left(-\beta \frac{
\hat{A}(q_i;\vec{e}) + \hat{A}(q_j;\vec{e})}{2} \right) \cr
&=&
\mathcal{S}_{ij}
\sqrt{\exp\left(-\beta (
\hat{A}(q_j;\vec{e}) + \hat{A}(q_j;\vec{e})) \right)} \cr
&=&
\mathcal{S}_{ij}
\sqrt{
\hat{\pi}(q_i;\vec{e}) 
\hat{\pi}(q_j;\vec{e})} \, .
\label{eq:geomAvg}
\end{eqnarray}
Eq.~\ref{eq:geomAvg} expresses the stationary distribution along the intersecting surface as geometric average between the stationary distributions of the cells.

Applying the assumption of small cells to the term $\nicefrac{1}{\hat{\pi}_i}$ in front of eq.~\ref{eq:GalerkinDiscretization}, we obtain
\begin{eqnarray}
\hat{\pi}_i 
&=& 
\int_{\Omega_i}
\hat{\pi}(q;\vec{e}) \, \mathrm{d}q \cr
&\approx&
\hat{\pi}(q_i;\vec{e}) \mathcal{V}_i \,,
\label{eq:pii}
\end{eqnarray}
where $\mathcal{V}_i$ is the volume of the $i$th subset.

Combining eqs.~\ref{eq:flux}, \ref{eq:geomAvg}, \ref{eq:pii} into eq.~\ref{eq:GalerkinDiscretization2}, we get the following expression for rates between adjacent cells:
\begin{eqnarray}
\hat{Q}_{ij,\,\mathrm{adj}}(\vec{e})
&=& 
\frac{\hat{D}(q_i;\vec{e}) + \hat{D}(q_j;\vec{e})}{2}
\,
\frac{\mathcal{S}_{ij}}{d_{ij} \, \mathcal{V}_i} \,
\sqrt{\frac{
\hat{\pi}(q_j;\vec{e})}{
\hat{\pi}(q_i;\vec{e})}}  \, ,
\label{eq:rate_adjacent}
\end{eqnarray}
which fulfill the matrix $\mathbf{\hat{Q}}$:
\begin{eqnarray}
\hat{Q}_{ij}(\vec{e}) &=& 
\begin{cases}
\hat{Q}_{ij,\,\mathrm{adj}}(\vec{e})
&\mbox{if  $i\ne j$, and  $\Omega_i$ is adjacent to $\Omega_j$}  \\
0                                           &\mbox{if  $i\ne j$, and  $\Omega_i$ is not adjacent to $\Omega_j$} \\
-\sum_{j=1, j\ne i}^K \hat{Q}_{ij}(\vec{e})                            &\mbox{if } i=j \, .
\end{cases} 
\label{eq:RateMatrix}
\end{eqnarray}

We remark that the grand canonical partition function (eq.~\ref{eq:GrandCanPartFunc4}) which appears in eq.~\ref{eq:ProbRelCoord}, cancels in the ratio in eq.~\ref{eq:rate_adjacent}, therefore, it is not necessary to calculate it.

\subsubsection{Approximating the grand canonical distribution}
In applications, the distribution $\hat{\pi}(q;\vec{e})$ of the grand canonical ensemble in eq.~\ref{eq:rate_adjacent} is approximated by its histogram, formally defined as
 \begin{eqnarray}
 \hat{h}(q;\vec{e}) 
 &=&
 \sum_{i=1}^K 
 \frac{
 \langle \hat{\pi}(q;\vec{e}) \vert \, \mathbf{1}_i(q) \rangle_{\hat{\pi}}
 }
 {
\langle 
\mathbf{1}_i(q)
\vert \, \mathbf{1}_i(q) 
\rangle_{\hat{\pi}}
}\cdot\,  \mathbf{1}_i(q) \, .
 \end{eqnarray}
Accordingly to the grand canonical distribution definition, the histogram $\hat{h}(q;\vec{e})$ can be estimated as weighted average:
\begin{eqnarray}
\hat{h}(q;\vec{e}) = \sum_{n=1}^s w_n(\vec{e}) \hat{h}_n(q;\vec{e}) \, ,
\label{eq:hist2}
\end{eqnarray}
where the canonical histograms $\hat{h}_n(q;\vec{e})$ are obtained from the trajectories of the scenarios.
Note that, since the state space $\Omega$ is partitioned by combining the simulations of the scenarios together, the histogram $\hat{h}_n$ of a single scenario contains entries with 0 value.
Therefore, to avoid empty entries also in the grand canonical histogram $\hat{h}(q;\vec{e})$ and \textsf{NaN} values in the SqRA rate matrix, a partial overlap of the histograms $\hat{h}_n$ is required, which, from a physical perspective, corresponds to overlapping of canonical ensembles of the scenarios.
This ensures that the entries of $\hat{h}(q;\vec{e})$ be strictly positive.

\subsubsection{Numerical methods to estimate the diffusion profile}
The calculation of the position-dependent diffusion is more complicated.
For high-dimensional systems, such as biomolecules, there are multiple numerical methods to estimate the diffusion profile from simulation data with different advantages and disadvantages.
Here, we briefly summarize the most used and reliable methods.

The simplest approach is to use the Stokes-Einstein law, which relates diffusion with the mean-square displacement of the positions of the atoms \cite{Frenkel2002}, however this method is limited to regions of the reaction coordinate where the derivative of the free energy profile is almost zero and the effective dynamics can be assumed pure Brownian \cite{Best2010, Lee2016}.
A more general solution exploits umbrella sampling simulations and estimates the position dependent diffusion from the Laplace transformation of the autocorrelation function of the velocities of the atoms along the reaction coordinate \cite{Woolf1994}.
This method can be simplified employing the autocorrelation function of the positions instead of velocities  \cite{Hummer2005}.
Within the framework of Bayesian analysis, the free energy profile and the diffusion profile can be determined from short trajectories  by optimizing a likelihood function \cite{Hummer2005}.
A different strategy employs the Smoluchowski equation to get the position dependent diffusion from the round trip time, i.e. the sum of the mean first passage times to cross the barrier between two adjacent metastable states in both the directions \cite{Hinczewski2010,Sedlmeier2011}.  
The most recent proposed method builds on Markov state model analysis and the Kramers-Moyal expansion, and by means of the dynamic histogram analysis method (DHAM) \cite{Rosta2015}, can be applied to both unbiased and biased simulations \cite{Sicard2021}.

In application, depending on the system under investigation, one of these methods can be used to estimate the canonical diffusion profiles $\hat{D}_n(q)$ of the $s$ scenarios.
The grand canonical diffusion for specific environmental coordinates introduced in eq.~\ref{eq:FP2}, similarly to what we have done for the grand canonical distribution, is then estimated as a weighted average:
\begin{eqnarray}
\hat{D}(q;\vec{e}) = \sum_{n=1}^s w_n(\vec{e})^2\hat{D}_n(q)  \, .
\label{eq:De}
\end{eqnarray}
In eq.~\ref{eq:De} the weights are squared because, from a mathematical point of view, diffusion coefficients are interpreted as variances.

\subsubsection{Discretization error}
A disadvantage of the proposed strategy is that the crisp discretization of the state-space induces a discretization error 
\begin{eqnarray}
\delta = \Vert 
\hat{\pi}(q;\vec{e}) - \hat{h}(q;\vec{e})
\Vert_{\hat{\pi}} \, ,
\end{eqnarray}
with $\Vert 
\cdot
\Vert_{\hat{\pi}}
=
\langle \cdot \vert \, \cdot \rangle_{\hat{\pi}}^{\nicefrac{1}{2}}
$, which is propagated by the SqRA to the solution $\hat{\rho}(q,t;\vec{e})$ of the Fokker-Planck equation and to the associated dynamic observables, such as rates.

Recently, it has been proved that, for a Voronoi tessellation whose distance between adjacent cells $d_{ij}$ is bounded from above, i.e. $d_{ij} < C d$ for some constant $C>0$, then SqRA and similar discretization schemes have a quadratic convergence in $d$ \cite{Heida2021}.
Additionally, if $d \rightarrow 0$, the SqRA converges to the Fokker-Planck equation \cite{Donati2018b, Donati2021, Heida2018}. 
As a consequence, one may arbitrarily reduce the discretization error by increasing the number of cells.
However, the reduction of the discretization error leads to an increase in statistical uncertainty, because the average number of counts per cell is lower.
For a rigorous analysis of the problem, we refer to the work of M.~Heida et al.~\cite{Heida2021}.

A second source of error is related to the calculation of the diffusion.
The described methods require the use of time correlation functions or transition counting matrices which depend on a lag-time.
It is important, however, to verify whether the choice of the lag-time meets the requirement of Markovianity, since the underlying assumption is that the dynamics of the system in the reduced space can be accurately described by a diffusion model.
For this purpose, several tests are available \cite{Best2011}.
For example, one can estimate the diffusion profile $\hat{D}(q;\vec{e},\tau)$, and consequently the rate matrix $\hat{\mathbf{Q}}(\vec{e},\tau)$, at different lag times.
If the discretization error is negligible, then the eigenvalues and the eigenvectors quickly converge to a constant function, i.e. they do not depend on the choice of the lag-time \cite{Prinz2011,keller2019}.
\subsection{Robust Perron Cluster Cluster Analysis}
\label{subsec:pcca}
The robust Perron Cluster Cluster Analysis (PCCA+) algorithm \cite{Deuflhard2004, Kube2007, Roblitz2013, Weber2018, Erlekam2019} is used to cluster the $K$ Voronoi cells into $n_c$ macrostates and to cluster the $K\times K$ rate matrix $\hat{\mathbf{Q}}(\vec{e})$ into a $n_c \times n_c$ rate matrix $\hat{\mathbf{Q}}_c(\vec{e})$ of the macrostates.
To do so, first we need to compute the eigenvectors of the matrix $\hat{\mathbf{Q}}(\vec{e})$, which satisfy the equation
\begin{eqnarray}
\hat{\mathbf{Q}}(\vec{e}) \hat{X}  = \Lambda \hat{X}  \, ,
\end{eqnarray}
where $\hat{X}$ is a $K\times K$ matrix with the right-eigenvectors $\lbrace \hat{X}_1,...,\hat{X}_K \rbrace$ as columns, and $\Lambda$ is the diagonal matrix with eigenvalues $(\lambda_1,...,\lambda_K)$ as diagonal components.
The first eigenvector $\hat{X}_1$ is constant with positive entries and eigenvalue $\lambda_1=0$.
The other eigenvectors have positive and negative entries with eigenvalues strictly negative: $\lambda_i<0$ $\forall i>1$.
In particular, we distinguish between the first $n_c-1$ dominant eigenvectors with eigenvalues $\lambda_2,...,\lambda_{n_c}$ near 0 which represent the slow processes of the dynamical systems, and all the other eigenvectors which are associated to the fastest processes.

The dominant eigenvectors $\lbrace \hat{X}_2,...,\hat{X}_{n_c}\rbrace$ constitute a $K \times (n_c-1)$ matrix $\hat{X}'$ whose rows are the vertices of a $(n_c-1)$-simplex. 
The simplex has a physical interpretation: the vertices represent the conformations of the system, the points on the edges represent the transition regions.

PCCA+ transforms the matrix $\hat{X}'$ of the dominant eigenvectors into a $K\times n_c$ matrix $\hat{\chi}$ whose rows are the vertices of a $(n_c-1)$-standard simplex according to
\begin{eqnarray}
\hat{\chi}= \hat{X}'\mathcal{A} \, ,
\end{eqnarray}
where we have introduced the transformation matrix $\mathcal{A}$.
The sum of the rows of the matrix $\hat{\chi}$ is equal to 1, then we interpret the columns of the new matrix $\hat{\chi}$ as membership functions which indicate the probability that a microstate belongs to a certain macrostate.
Intuitively, we assert that a microstate $q$ with $\hat{\chi}_i(q) \approx 1$ belongs to the $i$th conformation, thus PCCA+ is first of all a technique to identify macrostates as fuzzy sets.

Finally, given the membership functions $\hat{\chi}$ and the stationary distribution $\hat{\pi}$, the rate matrix $\hat{\mathbf{Q}}(\vec{e})$ is reduced to the $n_c\times n_c$ matrix of the conformations 
\begin{eqnarray}
\hat{\mathbf{Q}}_c(\vec{e})
&=&
(\hat{\chi}^\top \mathrm{diag}(\hat{\pi}) \hat{\chi})^{-1} \hat{\chi}^\top \mathrm{diag}(\hat{\pi}) \hat{\mathbf{Q}}(\vec{e}) \hat{\chi}  \, ,
\label{eq:Qc}
\end{eqnarray}
which expresses the transition rates between fuzzy sets.
\subsection{Simulations at constant environmental variables}
\label{subsec:constpH}
We generalized the method ``Hybrid nonequilibrium Molecular Dynamics-Monte Carlo simulation method'' (neMDMC) proposed by Y. Chen et al.~\cite{Chen2015} and based on the work of H. Stern~\cite{Stern2007}, to run simulations of the GCE at given environmental variables $\vec{e}$.
The method generates trajectories by repeating MD simulations of length $\tau_{\mathrm{MD}}$ of the single scenarios, followed by non-equilibrium MD (neMD) simulations of length $\tau_{\mathrm{ne}}$ to switch scenario.
The non-equilibrium simulations between scenarios are accepted or rejected according to a Metropolis criterion. 

This approach requires to define the potential which governs the transition between scenarios
\begin{eqnarray}
U_{\mathrm{ne}}(\mathbf{r},\lambda(t_{\mathrm{ne}})) 
&=& 
\lambda(t_{\mathrm{ne}})U_n(\mathbf{r}) + \left[1 - \lambda(t_{\mathrm{ne}})\right] U_m(\mathbf{r}) \, ,
\label{eq:pHpot}
\end{eqnarray}
where $\lambda$ takes the value of 1 if the system activates the $n$th scenario, and 0 if the system activates the $m$th scenario.
The parameter $\lambda$ evolves with the time variable $t_{\mathrm{ne}}$ of the non-equilibrium trajectory as
\begin{eqnarray}
\lambda(t_{\mathrm{ne}}) = 1 - \frac{t_{\mathrm{ne}}}{\tau_{\mathrm{ne}}} \, .
\end{eqnarray}
To accept or reject a transition, we define the Metropolis criterion
\begin{eqnarray}
&\mathcal{T}(U_n(\mathrm{r}) \rightarrow U_m(\mathrm{r})) =
\min\left\lbrace 1, \exp\left[-\beta \left(U_m(\mathrm{r}) + \gamma(\vec{e}) - U_n(\mathrm{r}) \right)\right]
\right\rbrace \, ,
\label{eq:metropolis}
\end{eqnarray}
where we have introduced the $\vec{e}$-dependent constant
\begin{eqnarray}
\gamma(\vec{e}) = -\frac{1}{\beta} \log \frac{w_n(\vec{e})}{w_m(\vec{e})} - \Delta G \, ,
\label{eq:gamma}
\end{eqnarray}
needed to satisfies the ratio between the statistical weights of the scenarios (eq.~\ref{eq:ScenarioWeight}).
In eq.~\ref{eq:gamma}, $\Delta G$ is the free energy
\begin{eqnarray}
\Delta G = 
- \frac{1}{\beta}
\log
\frac{
\int \mathrm{d} \mathbf{r} \, e^{-\beta U_n(\mathbf{r})}
}{
\int \mathrm{d} \mathbf{r} \, e^{-\beta U_m(\mathbf{r})}
}\, .
\label{eq:freeEnergy}
\end{eqnarray}
The method produces trajectories that explore the whole GCE and which can be used to estimate the grand canonical distribution on relevant coordinates (eq.~\ref{eq:ProbRelCoord}).
\newpage
\section{Results}
\label{sec:results}
\subsection{Numerical experiment}
\begin{figure}[ht]
    \centering
    \includegraphics[scale=1]{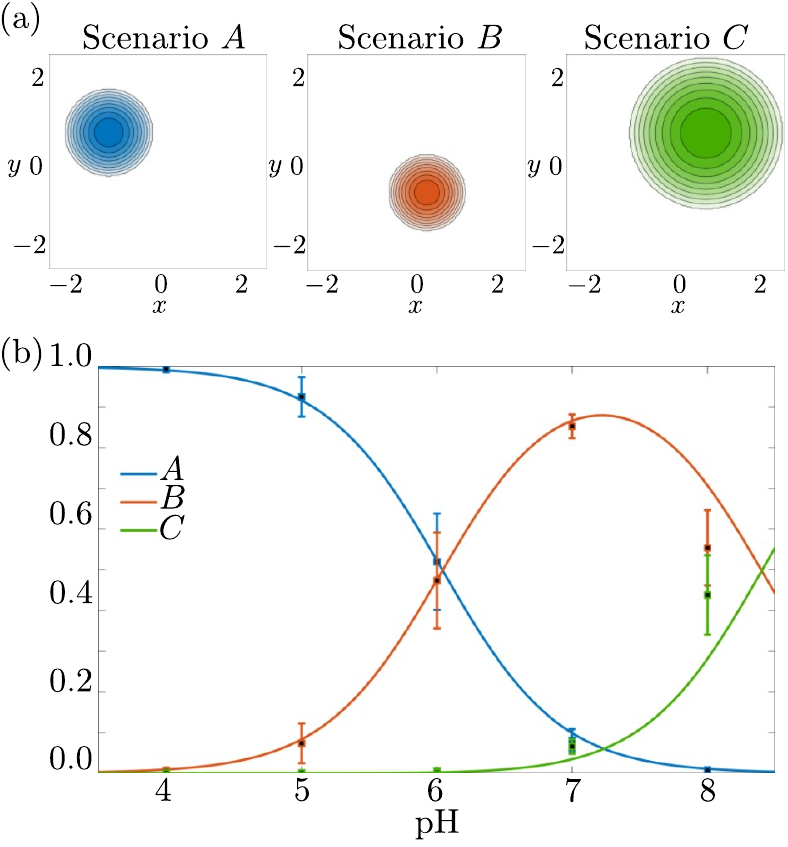}
    \caption
    {
    (a) Potential energy functions of the three scenarios; (b) Probability of occurrence of each scenario: the solid lines represent the analytical functions (eqs.~\ref{eq:HHeq1}, ~\ref{eq:HHeq2}, ~\ref{eq:HHeq3}), the  squares with the error bars were estimated by constant-pH simulations. 
    } 
    \label{fig:fig1}
\end{figure}

We studied and applied the GCEkinCEs procedure to a two-dimensional system defined by the following potentials:
\begin{eqnarray}
V_{A}(x,y) &=& \frac{15}{2} \left(x + \frac{3}{2} \right)^2 + \frac{15}{2} \left(y - \frac{1}{2}\right)^2\, , \cr
V_{B}(x,y) &=& \frac{20}{2} \left(x - \frac{1}{2} \right)^2 + \frac{20}{2} \left(y + \frac{3}{2}\right)^2\, , \cr
V_{C}(x,y) &=& \frac{5}{2} \left(x - \frac{1}{2}\right)^2 + \frac{5}{2} \left(y - \frac{1}{2}\right)^2\, .
\label{eq:potentials}
\end{eqnarray}
The three potentials, plotted in fig.~\ref{fig:fig1}-a, represent the three scenarios of the ligand-receptor system described in ref.~\cite{Ray2020}:
\begin{enumerate}
    \item Scenario $A$: both the ligand and the receptor are protonated. 
    The probability of occurrence of scenario $A$ is the joint probability that both the receptor and the ligand are protonated:
    \begin{eqnarray}
    w_{A}(\mathrm{pH}) &=& 
    \frac{1}{10^{\mathrm{pH} - \mathrm{p}K_{a1}} + 1}
    \cdot
    \frac{1}{10^{\mathrm{pH} - \mathrm{p}K_{a2}} + 1} \, ,
    \label{eq:HHeq1}
    \end{eqnarray}
    where $\mathrm{p}K_{a1}$ and $\mathrm{p}K_{a2}$ are respectively the $\mathrm{p}K_{a}$ values of the receptor and the ligand.
    The derivation is reported in Appendix \ref{sec:App1}.
    \item Scenario $B$: Only the ligand is protonated. 
    The probability of occurrence of scenario $B$ is obtained by multiplying the probability that the receptor is deprotonated by the probability that the ligand is protonated:
    \begin{eqnarray}
    w_{B}(\mathrm{pH}) &=& \frac{1 - w_{A}(\mathrm{pH})}{10^{\mathrm{pH} - \mathrm{p}K_{a2}} + 1} 
    \label{eq:HHeq2}
    \end{eqnarray} 
    \item Scenario $C$: neither the ligand, nor the receptor are protonated. 
    The probability of occurrence of scenario $C$ is derived from the partition of unity defined in eq.~\ref{eq:WeightsCondition}:
    \begin{eqnarray}
    w_{C}(\mathrm{pH}) &=&1 - w_{A}(\mathrm{pH}) - w_{B}(\mathrm{pH})
    \label{eq:HHeq3}
    \end{eqnarray}
\end{enumerate}
%
%
In our numerical experiment we used the histidine $\mathrm{p}K_{a1}$ value of 6.04 and the fentanyl $\mathrm{p}K_{a2}$ value of 8.4 (ref.~\cite{Ray2020}).
The weight functions for these $\mathrm{p}K_a$ values are plotted in fig.~\ref{fig:fig1}-b.
Note that the $\mathrm{p}K_a$ difference is such that eq.~\ref{eq:HHeq1} can be approximated as
\begin{eqnarray}
w_{A}(\mathrm{pH}) &=& \frac{1}{10^{\mathrm{pH} - \mathrm{p}K_{a1}} + 1}  \, .
\end{eqnarray}

We set the Boltzmann constant $k_B = 8.31 \times 10^{-3} \,\mathrm{kJ\cdot mol^{-1}\cdot K^{-1}}$ and the temperature was $T=300\, \mathrm{K}$, then the thermodynamic beta was $\beta = \nicefrac{1}{k_B T}= 0.40 \, \mathrm{J^{-1}}$.
We assumed that the dynamics describes the motion of a particle of mass $m=1\, \mathrm{amu}$ and friction $\xi = 1\, \mathrm{ps}^{-1}$.
The diffusion constant was $D=\nicefrac{k_B T}{m \xi} = 2.49\,\mathrm{nm}^2\,\mathrm{ps}^{-1}$ and do not depend on the pH.
The Euler-Maruyama scheme was applied with an integrator timestep $\Delta t = 10^{-3} \, \mathrm{ps}$ and we generated three trajectories of $2\times10^5$ timesteps for each potential.
For the purpose of estimating statistical uncertainties, each simulation was independently replicated five times.

\begin{figure}[ht]
    \centering
    \includegraphics[scale=1]{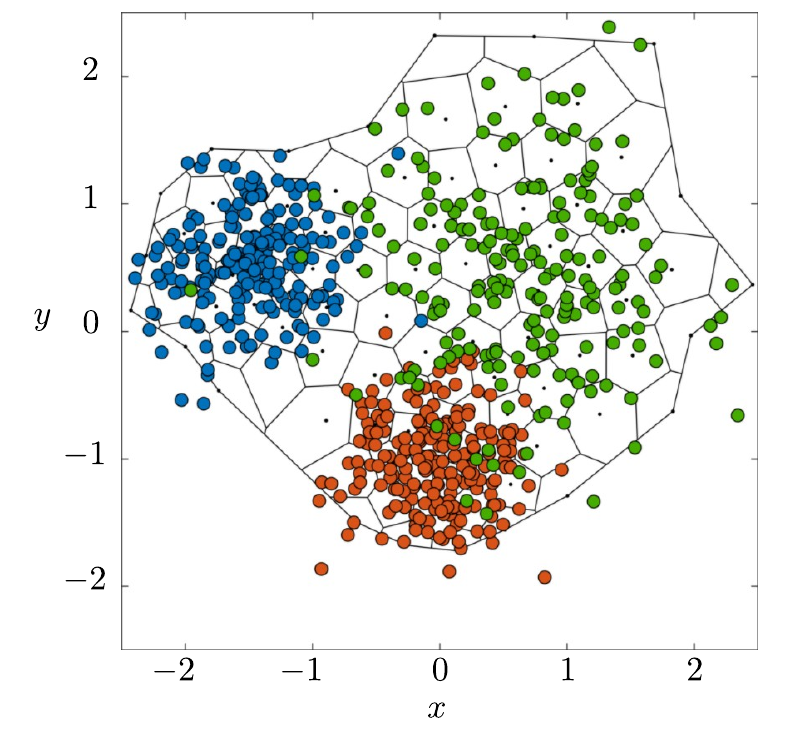}
    \caption
    {
    Scenarios trajectories generated independently.
    The sampled area is partitioned in 100 Voronoi cells.
    } 
    \label{fig:fig2}
\end{figure}

In fig.~\ref{fig:fig2}, 200 representative points from each trajectory are plotted, where the colors are used to identify the scenario.
The points of the scenario $C$ result to be more scattered due to the weaker spring constant which defines its potential.
Additionally, we observe that the trajectories cross each other, ensuring the overlap of the distributions required by our procedure to avoid numerical instabilities.
%
%

Next, we applied the $K$-means algorithm with $K=\, $25, 50, 100, 250, 500, and 1000, obtaining six different Voronoi tessellation of the space.
Fig.~\ref{fig:fig2} shows the Voronoi tessellation with 100 cells.
The algorithm was applied to a set of three  trajectories (one for each scenario) all together, and the boundaries were clipped to avoid Voronoi cells with an infinite volume.

\begin{figure*}[ht]
    \centering
    \includegraphics[width=\textwidth]{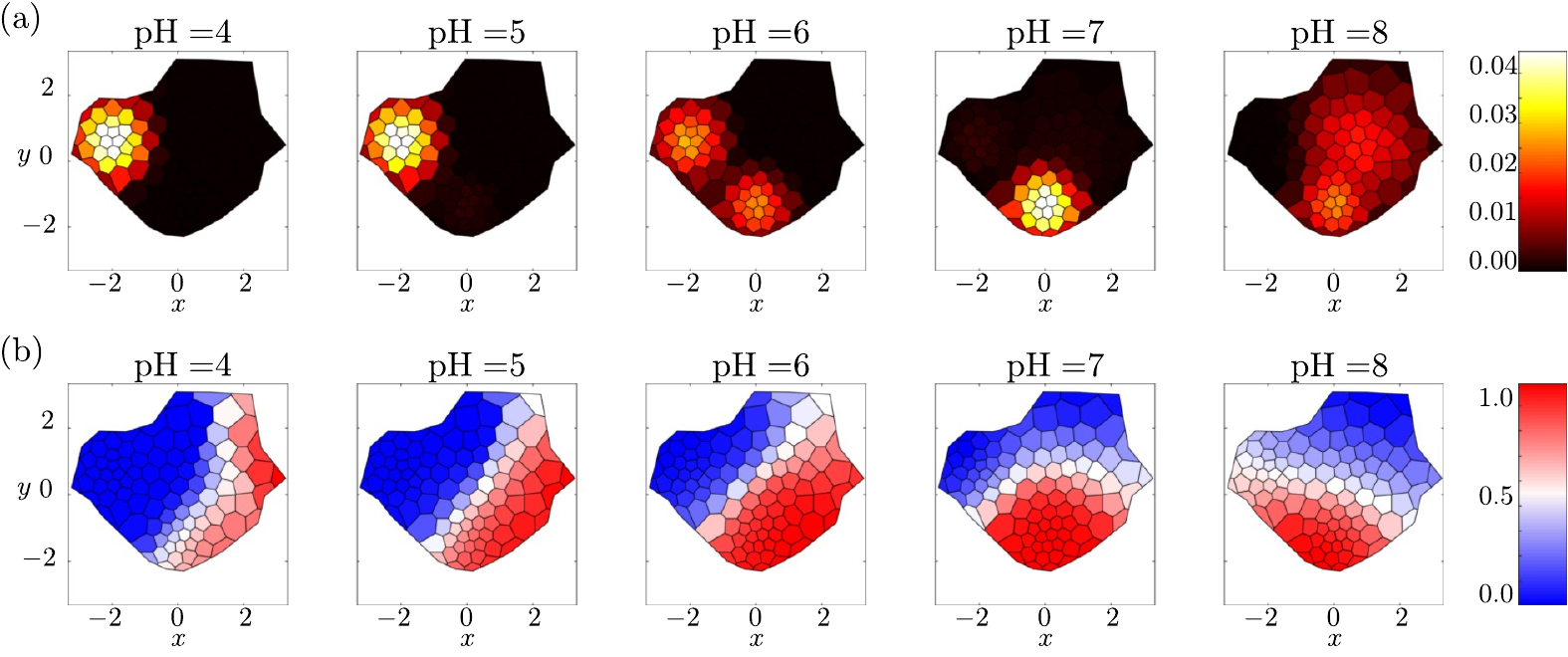}
    \caption
    {
    (a) Grand canonical stationary distributions for five different pH values obtained by GCEkinCEs; (b) Membership function $\chi_2(\mathrm{pH})$ of the second macrostate for five different pH values.
    The graphs refer to the 100-cells Voronoi discretization.
    } 
    \label{fig:fig3}
\end{figure*}

For each Voronoi discretization, we generated three histograms $h_A(x,y)$, $h_B(x,y)$ and $h_C(x,y)$ which approximate the canonical distribution of each scenario.
The three distributions were finally combined with the weights $w_A(\mathrm{pH})$, $w_B(\mathrm{pH})$ and $w_C(\mathrm{pH})$ (eqs.~\ref{eq:HHeq1}, ~\ref{eq:HHeq2}, ~\ref{eq:HHeq3}), according to eq.~\ref{eq:hist2}, to obtain an approximation of the grand canonical distribution for a given pH value.
The distributions for the 100-cells discretization, are plotted in fig.~\ref{fig:fig3}-a for five pH values between 4 and 8.
The dark colors represent a low, or very low probability that a Voronoi cell is occupied, conversely the light color represents a high probability of occupancy.
In agreement with the weights defined in eq.~\ref{eq:HHeq1}, ~\ref{eq:HHeq2}, ~\ref{eq:HHeq3}, at pH=4 the scenario $A$ is dominant.
At pH=6, the scenarios $A$ and $B$ have the same weights $w_A(\mathrm{pH})$ and $w_B(\mathrm{pH})$, thus we observe a distribution with two peaks of almost the same height.
The difference, not noticeable in this figure, is due to the different spring constant of the potentials $V_A(x,y)$ and $V_B(x,y)$.
At pH=8, the scenario $A$ disappears and the scenario $C$ becomes visible.
%
%

The SqRA rate matrix $\hat{\mathbf{Q}}(\mathrm{pH})$ (eq.~\ref{eq:RateMatrix}) was built for a range of pH values between 3.5 and 8.5, for each discretization.
Assuming the existence of two macrostates (bound and unbound system), we used PCCA+ to estimate the membership functions $\chi_1(\mathrm{pH})$ and $\chi_2(\mathrm{pH})$.
In fig.~\ref{fig:fig3}-b, we report the membership function $\chi_2$ which denotes the probability that a Voronoi cell belongs to the second macrostate.
Because $\chi_1$ is complimentary, in the sense that $\chi_1+\chi_2=1$, we interpret the red cells as belonging to the second macrostate, the blue cells as belonging to the first macrostate, and the white cells represent the transition region between macrostates.

The graphical representation well describes how the macrostates evolve and which is the direction of the dominant process.
At low pH, the first macrostate is dominant and it includes mainly the Voronoi cells associated to the scenario $A$.
At pH=6, the membership function is perfectly symmetric as the two macrostates are now defined by the scenarios $A$ and $B$ which have the same probability of occurrence.
At pH=7 the first macrostate begins to switch as scenario $A$ becomes less relevant. 
Finally, at pH=8 the first macrostate includes only cells of scenario $C$.

\begin{figure}[ht]
    \centering
    \includegraphics[scale=1]{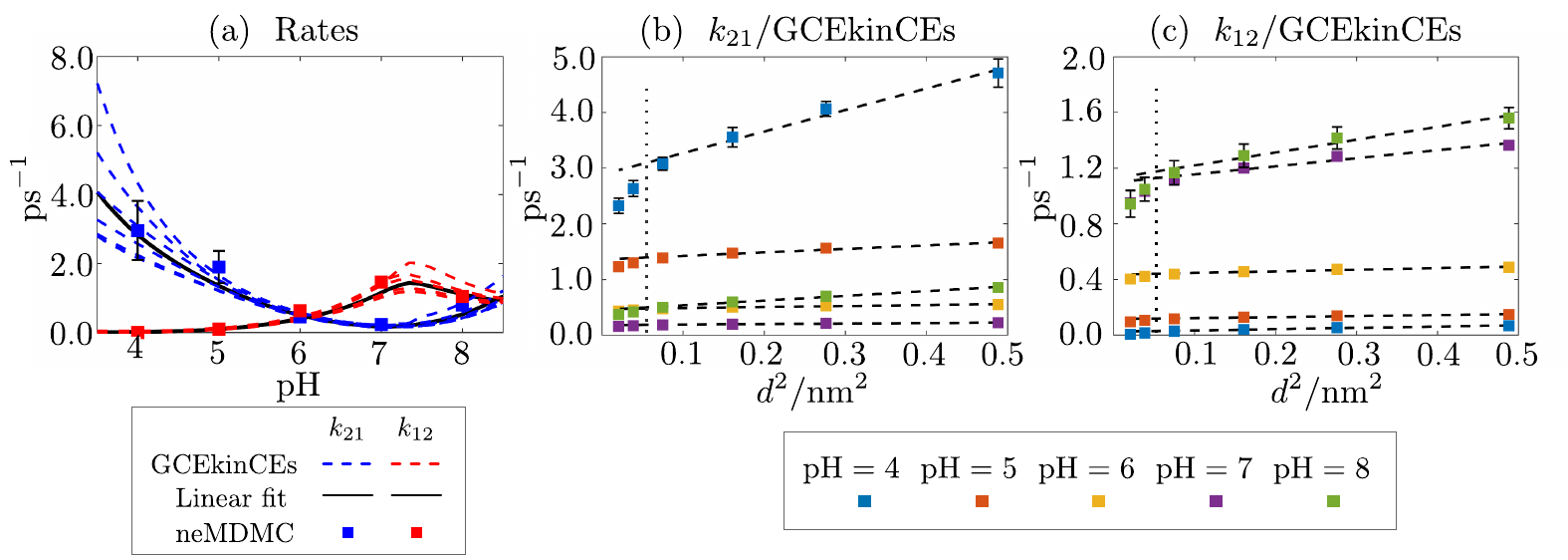}
    \caption
    {
    (a) 
    Transition rates $k_{21}$ (blue) and $k_{12}$ (red) as functions of the pH. 
    Each dashed line was obtained by GCEkinCEs with a different discretization: 25, 50, 100, 250, 500 and 1000 Voronoi cells, from top to bottom.
    The black solid lines represent the rates that one would obtain assuming infinitely small Voronoi cells ($d\rightarrow 0 \, \mathrm{nm}$) based on a linear fit of the rates.
    The linear fit was estimated from the rates using the Voronoi tessellation with 25, 50, 100, and 250 cells.
    The squares with the error bars represent the rates estimated by neMDMC simulations using the 100-cells discretization.
    (b,c) Transition rate $k_{21}$
    and $k_{12}$, estimated by GCEkinCEs, as functions of the square of the average distance $d$ between adjacent cells.
    Each square characterizes a pH value (color) and one of six the discretizations: 25, 50, 100, 250, 500 and 1000 Voronoi cells. 
    Each dashed line represents a linear fit, based on the four points at the right side of the dotted line.
    } 
    \label{fig:fig4}
\end{figure}

Applying eq.~\ref{eq:Qc}, we computed the rate matrix of the conformations
\begin{eqnarray}
\hat{\mathbf{Q}}_c(\mathrm{pH}) =
\begin{pmatrix}
k_{12}(\mathrm{pH}) & -k_{12}(\mathrm{pH})\\
k_{21}(\mathrm{pH}) & -k_{21}(\mathrm{pH})
\end{pmatrix} \, ,
\end{eqnarray}
where $k_{12}(\mathrm{pH})$ and $k_{21}(\mathrm{pH})$ are respectively the rates between the macrostates.
We repeated the same analysis for each discretization, and reported the results in fig.~\ref{fig:fig4}.

First, let us analyze how the rates are affected by discretization.
Fig.~\ref{fig:fig4}-a shows the rates $k_{12}$ (red) and $k_{21}$ (blue) as dashed lines estimated respectively with 25, 50, 100, 250, 500 and 1000 Voronoi cells, from the top to the bottom.
Upon first glance, it appears that the dashed curves converge as the number of cells increases.
Figs.~\ref{fig:fig4}-b,c show the dependence of the rates on the square of the average distance $d$ between each pair of adjacent cells: 
each $d$ value is associated respectively with a discretization of 1000, 500, 250, 100, 50, and 25 cells (from the left to the right), and each color represents a pH value as reported in the legend; the error bars have been estimated as standard deviation using the ensemble of replicas.
Only the four values on the right side of the dotted line (250, 100, 50, and 25 cells) show a linear dependence (black dashed line).
Instead, the two leftmost values (1000 and 500 cells), are underestimated with respect to the linear fit.
This observation indicates that, on the one hand, the SqRA is quadratic convergent with the square of the average distance between cell centers, as shown by M.~Heida \cite{Heida2021}; on the other hand, an excessive increase in the number of cells results in not negligible statistical errors.
In light of these considerations, we assume as optimal rates those extrapolated from the linear fit with $d = 0$.
These rates are represented in fig.~\ref{fig:fig4}-a as black solid lines.

Let's now analyze how the rates depend on the pH.
At low pH the system is trapped in the first macrostate which overlaps with scenario $A$.
Accordingly, the probability to leave the first macrostate is zero and $k_{12}(\mathrm{pH}) \approx 0$.
In contrast, if the system is configured in the second macrostate, it will evolve towards a configuration which belongs to the first macrostate and $k_{21}\approx 3:4\, \mathrm{ps}^{-1}$.
Increasing the pH, the system becomes bimetastable, i.e. it encompasses two quasi-stationary states, separated by a high barrier which prevents the transition.
It follows that $k_{21}$ drops to small values, as it is more difficult to reach the first macrostate from any other point of the state space, while $k_{12}$ slightly increases.
At pH = 6, we observe that $k_{12}\approx k_{21}\approx 0.5\, \mathrm{ps}^{-1}$, confirming the peculiar symmetry of the system at this pH value.
At pH=7 the system is still bimetastable, but we observe a sudden change of the rate $k_{12}$ due to the switch of scenarios: from $A+B$ to $B+C$.
%
%
\subsection{Method validation}
In order to validate our results, we used the constant-pH simulations method proposed by Y.~Chen \cite{Chen2015} and H.~ Stern \cite{Stern2007}.
Given a pH value, we generated a long trajectory alternating simulations of length $\tau_{\mathrm{MD}} = 5\times10^3$ timesteps performed in the active scenario, i.e. using one of the potentials defined in eq.~\ref{eq:potentials} from time to time; and non-equilibrium MD simulations of $\tau_{\mathrm{ne}}=50$ timesteps using the potential $U_{\mathrm{ne}}(\mathbf{r},\lambda(t_{\mathrm{ne}}))$, defined in eq.~\ref{eq:pHpot}, which interpolates the potentials between two scenarios.
The non-equilibrium simulations were accepted or rejected according to the Metropolis criterion defined in eq.~\ref{eq:metropolis} using the scenario probabilities given in eqs.~\ref{eq:HHeq1}, ~\ref{eq:HHeq2}, ~\ref{eq:HHeq3}.
The cycle MD-neMD was repeated 1000 times.
We used again the Euler-Maruyama scheme (eq.~\ref{eq:EMscheme}) with the integrator timestep $\Delta t = 10^{-3} \, \mathrm{ps}$.
In total, we carried out five independent simulations for five different pH values in the range [4:8], for a total amount of 25 trajectories, in order to estimate the uncertainty of rates for each pH value.

The constant $\gamma(\vec{e})$ defined in eq.~\ref{eq:gamma} was estimated using the statistical weights of the scenarios in eqs.~\ref{eq:HHeq1}, ~\ref{eq:HHeq2}, ~\ref{eq:HHeq3} and the integrals of the free energy differences in eq.~\ref{eq:freeEnergy} were computed by trapezoidal approximation.
This was possible because of the low dimensionality of the system under examination. 
Instead, high dimensional systems would require more advanced alchemical free energy calculation techniques.

\begin{figure*}[ht]
    \centering
    \includegraphics[width=\textwidth]{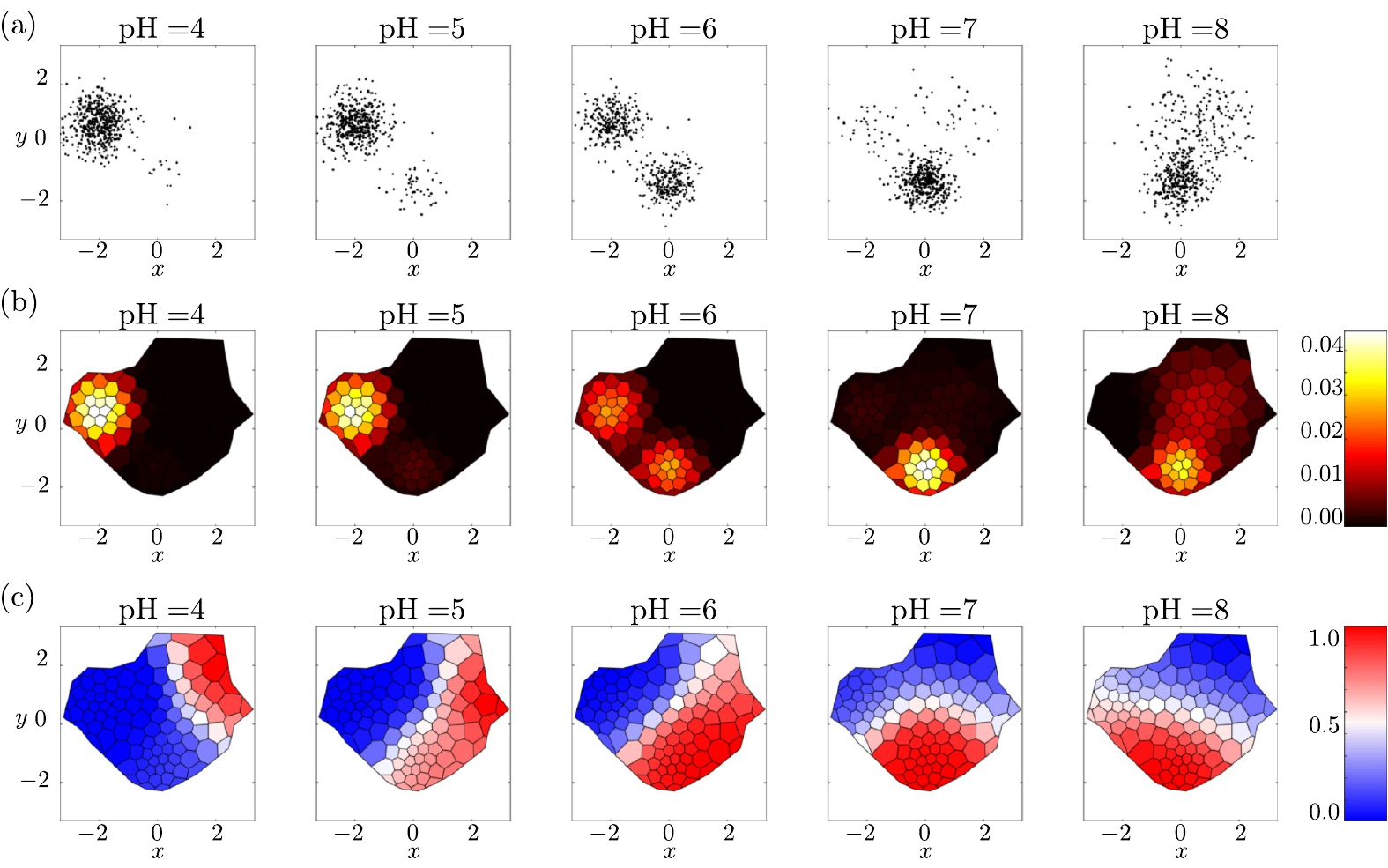}
    \caption
    {
    (a) Representative points of five trajectories generated at constant-pH by neMDMC; (b) Grand canonical stationary distributions for five different pH values estimated from the corresponding trajectories;
    (c) Membership function $\chi_2(\mathrm{pH})$ of the second macrostate for five different pH values.
    The graphs refer to the 100-cells Voronoi discretization.
    } 
    \label{fig:fig5}
\end{figure*}

Five representative trajectories are plotted in fig.~\ref{fig:fig5}-a.
Differently from the previous method, the trajectories do not sample separately the CEs of the scenarios (fig.~\ref{fig:fig3}), but the GCE for a given pH value.
Thus, increasing the pH, we observe how the trajectory initially samples the region occupied by the scenario $A$ (pH=4), then samples both the scenarios $A$ and $B$ (pH=6), and finally it samples the scenario $B$ and $C$ (pH=8).
We counted how many times each scenario was active during each simulation, and we estimated the activation frequencies as functions of the pH which result in good agreement with the probabilities of occurrence (eqs.~\ref{eq:HHeq1}, ~\ref{eq:HHeq2}, ~\ref{eq:HHeq3}, fig.~\ref{fig:fig1}).

The trajectories were used to calculate the grand canonical distributions, the rate matrix $\hat{\mathbf{Q}}(\mathrm{pH})$ via SqRA and the rate matrix of the conformations $\hat{\mathbf{Q}}_c(\mathrm{pH})$ via PCCA+.
Alternatively, instead of using the SqRA, one could build a MSM
\cite{Schuette1999b,  Deuflhard2000, Chodera2007, Buchete2008, Prinz2011, Keller2011, keller2019}, counting the transitions between Voronoi cells within a certain lag-time, and then applying the PCCA+.
In order to make the results comparable, we used the same Voronoi tessellation of 100 cells produced in the previous example.

The distributions and the membership functions are plotted in fig.~\ref{fig:fig3}, while the average rates accompanied by error bars are plotted in fig.~\ref{fig:fig4}. 
The results are in good agreement with those obtained by GCEkinCEs.
However, if pH=4 the membership function is differently oriented and the rate $k_{12}$ has an error bar significantly larger than the other cases.
We suspect that this is an artifact resulting from the non-overlapping of the trajectory and the Voronoi tessellation.
Indeed, the trajectory mostly covers the regions of scenario $A$ and $B$, leaving the tessellation cells located in scenario $C$ empty.
Furthermore, the sampled points in zone $B$ are too few and the PCCA+ algorithm does not capture the presence of a barrier between the scenarios $A$ and $B$, but recognizes the two regions as a unique metastable state.
The artifact is eliminated at pH = 5, the barrier between $A$ and $B$ is lower, and the trajectory is better able to sample both regions, allowing for PCCA+ to identify the bi-metastability.
%
%
%
%
%
%
%
\section{Conclusion and outlook}
\label{sec:conclusion}
Modelling the influence of different pH values on molecular systems requires changing the protonation state of specific molecular moieties according to a probability which depends on the $\mathrm{p}K_a$ value of that group and on the pH value of the environment. 

There are two different extreme ways to account for this probability in a molecular ensemble setting. 
In the first strategy, one assumes that the protonation states of each entity of the ensemble changes rapidly enough in the course of time, such that a single trajectory ``running through'' the conformational and protonation states of the system, captures the correct GCE distribution. 
In hybrid MDMC simulations \cite{Burgi2002, Mongan2004,Williams2010,Baptista2002,Meng2010,Dashti2012,Swails2014,Stern2007, Chen2015, Radak2017}, the changing rate of protonation states does not stem from physical rate measurements, but is adjusted such that the mixing properties of the ensemble simulation are statistically convenient. 
Then, the rates of molecular conformational changes, estimated from the generated trajectory, depend on how the changing of the protonation states is algorithmically organized. 

In our new approach, called GCEkinCEs, we model the other extreme point of view. 
We assume that the protonation states are distributed according to the molecular and environmental parameters, and we reconstruct the GCE distribution from separate simulations of the CEs of the protonation states.
The rates of molecular conformational changes are, therefore, not to be taken from the simulation of single trajectories as in MDMC approaches, but directly from the distribution of the grand canonical ensemble.
This is possible thanks to SqRA \cite{Lie2013, Donati2018b, Heida2018, Donati2021} and PCCA+ \cite{Deuflhard2004, Kube2007, Roblitz2013, Weber2018, Erlekam2019}, used respectively to build the fine-grained rate matrix  and the coarse-grained rate matrix of the conformations.
One of the main advantages of GCEkinCEs is the efficiency of the pH-reweighting, which allows for generating continuous functions representing the dependence between conformational transition rates and environmental pH. 

Furthermore, GCEkinCEs provides an important insight into the dynamics of conformations.
The analysis of the membership functions shows that the location of the transition region between two molecular conformational states on the free energy landscape depends on the pH value.
In other words, the environmental pH determines what sort of molecular motion is happening in the molecular system on the longest timescales.
Thus, this continuous pH-dependency is characterized mathematically via projection methods.

GCEkinCEs can be easily generalized by taking into account more than one environmental factors, however, extensive use would require a priori knowledge of the weights of the scenarios.
This remains a challenge that needs to be addressed, which, however, is strictly related to the nature of the variables under investigation.
In this work we used the Henderson-Hasselbach equation to estimate the protonation probability, but this approach is limited to systems with few titratable sites.
For systems with multiple titratable sites such as proteins, the protonation probabilities need a more complex description which takes into account also the pH-dependence of the interactions between titratable sites.
This requires the calculation of pH-dependent free energies as described in Appendix \ref{sec:App1} and the use of free energy calculation techniques.
In this regard, a new algorithm based on $\lambda$-dynamics \cite{Kong1996} has been recently implemented in GROMACS and promises a rapid calculation of titration curves, used to estimate the protonation probabilities, also for large systems with multiple protonation states \cite{Aho2022}.
Likewise, a similar strategy could be implemented for other environmental variables such as the salt or ion concentration which are responsible for several biological functions of biomolecules such as photosynthesis, respiration, metabolism, and signaling processes \cite{Prabhulkar2012,Shao2022,Srivastava2020,Chen2022}.
The related equilibrium equations do indeed lead to Henderson-Hasselbach-like equations which can be used to compute free energy differences, and the associated probabilities, as functions which depend on the concentration of salt or ions \cite{Ullmann1999}.
In more advanced models, one could also study the dependence on parameters characterizing the cell' shape, such as the membrane curvature which is suspected to play a role on protein kinetics \cite{Busch2015,Zeno2019,Rui2022}.

An interesting perspective and possible direction for further development of our method, is to redefine the problem as an optimization problem.
The weights could be determined by minimizing the statistical error, in a manner similar to that used for the WHAM equations, in which the stationary distribution is obtained as weighted average \cite{Ferrenberg1989,Kumar1992,Kastner2011}.

In conclusion, considering its versatility, GCEkinCEs has a considerable potential.
We believe it could be useful in drug design in combination with virtual screening applications \cite{Gorgulla2020}.
It may indeed contribute to the refinement of the selection of candidate ligands, taking into account characteristics of the cellular environment that are known to be related to the health of tissue.
%
%
\section*{DATA AVAILABILITY STATEMENT}
The data that support the findings of this study are available from the corresponding author upon reasonable request.
\begin{acknowledgments}
This research has been funded by the Deutsche Forschungsgemeinschaft (DFG, German Research Foundation) Cluster of Excellence MATH+, project AA1-15: ``Math-powered drug-design''.
\end{acknowledgments}
%

\appendix
\section{Derivation of protonation probabilities}
\label{sec:App1}
Here, we derive the functions defined in eqs.~\ref{eq:HHeq1}, \ref{eq:HHeq2} and \ref{eq:HHeq3} used to calculate the probability of occurrence in the numerical experiment.
Consider the protonation equilibrium formalized by the equation
\begin{eqnarray}
    \mathrm{HA} \rightleftharpoons  \mathrm{A}^- + \mathrm{H}^+ \, ,
\end{eqnarray}
where $\mathrm{HA}$ represents the protonated state, $\mathrm{A}^-$ is the deprotonated state and $\mathrm{H}^+$ is a hydrogen ion.
The acid dissociation constant (equilibrium constant) is 
\begin{eqnarray}
    K_a = 
    \frac{\left[\mathrm{A}^-\right] \left[\mathrm{H}^+\right]}
    {\left[\mathrm{HA}\right]} \, .
    \label{eq:eqConst}
\end{eqnarray}
We define the pH of the solution and the p$Ka$ value of an acid respectively as
\begin{eqnarray}
    \mathrm{pH} = - \lg \left[\mathrm{H}^+\right] \, , 
    \label{eq:pH}
\end{eqnarray}
and
\begin{eqnarray}
    \mathrm{p}K_a = - \lg K_a \, , 
    \label{eq:pKa}
\end{eqnarray}
where $\lg$ denotes the logarithm with base 10.
Inserting eqs.~\ref{eq:pH}, \ref{eq:pKa} into the decadic logarithm of eq.~\ref{eq:eqConst} yields the Henderson-Hasselbalch equation
\begin{eqnarray}
    \mathrm{pH} = \mathrm{p}K_a + \lg \frac{\left[\mathrm{A}^-\right]}{\left[\mathrm{HA}\right]} \, .
    \label{eq:HH}
\end{eqnarray}
At a given pH value, the probability that the acid $\mathrm{HA}$ is protonated is 
\begin{eqnarray}
p_{\mathrm{HA}} = \frac{\left[\mathrm{HA}\right]}{\left[\mathrm{HA}\right] + \left[\mathrm{A}^-\right]} \, ,
\label{eq:pHA}
\end{eqnarray}
and correspondingly the probability of the acid to be in the deprotonated state is 
\begin{eqnarray}
p_{\mathrm{A}} = 1 - p_{\mathrm{HA}} \, .
\end{eqnarray}
Rearranging eqs.~\ref{eq:pHA} and \ref{eq:HH} yields
\begin{eqnarray}
p_{\mathrm{HA}} = 
\frac{10^{\mathrm{p}K_a - \mathrm{pH}}}
{1 + 10^{\mathrm{p}K_a - \mathrm{pH}}} \, ,
\end{eqnarray}
or equivalently
\begin{eqnarray}
p_{\mathrm{HA}} = 
\frac{1}
{1 + 10^{\mathrm{pH}- \mathrm{p}K_a}} \, .
\end{eqnarray}

For more complex systems with $N$ titratable states (e.g. proteins), things get more involved as the Henderson-Hasselbalch equation does not take into account the interaction between titratable groups which is also pH-dependent.
In such a situation, the protonation probability of each residue can be estimated as thermodynamic average over all possible protonation states of the protein.
Following ref.~\cite{Ullmann1999}, we introduce the vector $\vec{s} = \lbrace s_1, s_2, ..., s_N \rbrace$ where $N$ denotes the number of titratable groups and each components $s_\mu$ adopts either the value 1 or 0 depending on whether group $\mu$ is protonated or deprotonated.
As there are $2^N$ protonation states, in order to average over all the cases, the protonation probability of the residue $\mu$ is written as
\begin{eqnarray}
w_\mu(\mathrm{pH}) &=& 
\frac
{
\sum_{i=1}^{2^N} s_\mu^i \exp\left(-\beta G^i\right)
}
{
\sum_{i=1}^{2^N} \exp\left(-\beta G^i\right) \, , 
}
\end{eqnarray}
where $\beta$ is the thermodynamic beta, and $G^i$ is the free energy of the $i$th protonation state, i.e. the energy required to protonate the $i$th group at a given pH and temperature.
%

\nocite{*}

\begin{thebibliography}{88}%
\makeatletter
\providecommand \@ifxundefined [1]{%
 \@ifx{#1\undefined}
}%
\providecommand \@ifnum [1]{%
 \ifnum #1\expandafter \@firstoftwo
 \else \expandafter \@secondoftwo
 \fi
}%
\providecommand \@ifx [1]{%
 \ifx #1\expandafter \@firstoftwo
 \else \expandafter \@secondoftwo
 \fi
}%
\providecommand \natexlab [1]{#1}%
\providecommand \enquote  [1]{``#1''}%
\providecommand \bibnamefont  [1]{#1}%
\providecommand \bibfnamefont [1]{#1}%
\providecommand \citenamefont [1]{#1}%
\providecommand \href@noop [0]{\@secondoftwo}%
\providecommand \href [0]{\begingroup \@sanitize@url \@href}%
\providecommand \@href[1]{\@@startlink{#1}\@@href}%
\providecommand \@@href[1]{\endgroup#1\@@endlink}%
\providecommand \@sanitize@url [0]{\catcode `\\12\catcode `\$12\catcode
  `\&12\catcode `\#12\catcode `\^12\catcode `\_12\catcode `\%12\relax}%
\providecommand \@@startlink[1]{}%
\providecommand \@@endlink[0]{}%
\providecommand \url  [0]{\begingroup\@sanitize@url \@url }%
\providecommand \@url [1]{\endgroup\@href {#1}{\urlprefix }}%
\providecommand \urlprefix  [0]{URL }%
\providecommand \Eprint [0]{\href }%
\providecommand \doibase [0]{http://dx.doi.org/}%
\providecommand \selectlanguage [0]{\@gobble}%
\providecommand \bibinfo  [0]{\@secondoftwo}%
\providecommand \bibfield  [0]{\@secondoftwo}%
\providecommand \translation [1]{[#1]}%
\providecommand \BibitemOpen [0]{}%
\providecommand \bibitemStop [0]{}%
\providecommand \bibitemNoStop [0]{.\EOS\space}%
\providecommand \EOS [0]{\spacefactor3000\relax}%
\providecommand \BibitemShut  [1]{\csname bibitem#1\endcsname}%
\let\auto@bib@innerbib\@empty
\bibitem [{\citenamefont {Haile}(1997)}]{Haile1997}%
  \BibitemOpen
  \bibfield  {author} {\bibinfo {author} {\bibfnamefont {J.~M.}\ \bibnamefont
  {Haile}},\ }\href@noop {} {\emph {\bibinfo {title} {{Molecular Dynamics
  Simulations: Elementary Methods}}}},\ Vol.\ \bibinfo {volume} {797 of Adv.
  Exp. Med. Biol.}\ (\bibinfo  {publisher} {Wiley-Interscience},\ \bibinfo
  {address} {New York},\ \bibinfo {year} {1997})\BibitemShut {NoStop}%
\bibitem [{\citenamefont {Frenkel}\ and\ \citenamefont
  {Smit}(2002)}]{Frenkel2002}%
  \BibitemOpen
  \bibfield  {author} {\bibinfo {author} {\bibfnamefont {D.}~\bibnamefont
  {Frenkel}}\ and\ \bibinfo {author} {\bibfnamefont {B.}~\bibnamefont {Smit}},\
  }\href@noop {} {\emph {\bibinfo {title} {{Understanding Molecular Simulation:
  From Algorithms to Applications}}}},\ \bibinfo {edition} {2nd}\ ed.\
  (\bibinfo  {publisher} {Academic Press},\ \bibinfo {address} {San Diego},\
  \bibinfo {year} {2002})\BibitemShut {NoStop}%
\bibitem [{\citenamefont {Manallack}\ \emph {et~al.}(2013)\citenamefont
  {Manallack}, \citenamefont {Prankerd}, \citenamefont {Yuriev}, \citenamefont
  {Oprea},\ and\ \citenamefont {Chalmers}}]{Manallack2013}%
  \BibitemOpen
  \bibfield  {author} {\bibinfo {author} {\bibfnamefont {D.~T.}\ \bibnamefont
  {Manallack}}, \bibinfo {author} {\bibfnamefont {R.~J.}\ \bibnamefont
  {Prankerd}}, \bibinfo {author} {\bibfnamefont {E.}~\bibnamefont {Yuriev}},
  \bibinfo {author} {\bibfnamefont {T.~I.}\ \bibnamefont {Oprea}}, \ and\
  \bibinfo {author} {\bibfnamefont {D.~K.}\ \bibnamefont {Chalmers}},\
  }\href@noop {} {\bibfield  {journal} {\bibinfo  {journal} {Chem. Soc. Rev.}\
  }\textbf {\bibinfo {volume} {42}},\ \bibinfo {pages} {485} (\bibinfo {year}
  {2013})}\BibitemShut {NoStop}%
\bibitem [{\citenamefont {Mongan}\ and\ \citenamefont
  {Case}(2005)}]{Mongan2005}%
  \BibitemOpen
  \bibfield  {author} {\bibinfo {author} {\bibfnamefont {J.}~\bibnamefont
  {Mongan}}\ and\ \bibinfo {author} {\bibfnamefont {D.}~\bibnamefont {Case}},\
  }\href@noop {} {\bibfield  {journal} {\bibinfo  {journal} {Curr. Opin.
  Struct. Biol.}\ }\textbf {\bibinfo {volume} {15}},\ \bibinfo {pages} {157}
  (\bibinfo {year} {2005})}\BibitemShut {NoStop}%
\bibitem [{\citenamefont {Chen}\ \emph {et~al.}(2014)\citenamefont {Chen},
  \citenamefont {Morrow}, \citenamefont {Shi},\ and\ \citenamefont
  {Shen}}]{Chen2014}%
  \BibitemOpen
  \bibfield  {author} {\bibinfo {author} {\bibfnamefont {W.}~\bibnamefont
  {Chen}}, \bibinfo {author} {\bibfnamefont {B.~H.}\ \bibnamefont {Morrow}},
  \bibinfo {author} {\bibfnamefont {C.}~\bibnamefont {Shi}}, \ and\ \bibinfo
  {author} {\bibfnamefont {J.~K.}\ \bibnamefont {Shen}},\ }\href@noop {}
  {\bibfield  {journal} {\bibinfo  {journal} {Mol. Simul.}\ }\textbf {\bibinfo
  {volume} {40}},\ \bibinfo {pages} {830} (\bibinfo {year} {2014})}\BibitemShut
  {NoStop}%
\bibitem [{\citenamefont {Barroso~da Silva}\ and\ \citenamefont
  {Dias}(2017)}]{Barroso2017}%
  \BibitemOpen
  \bibfield  {author} {\bibinfo {author} {\bibfnamefont {F.}~\bibnamefont
  {Barroso~da Silva}}\ and\ \bibinfo {author} {\bibfnamefont {L.}~\bibnamefont
  {Dias}},\ }\href@noop {} {\bibfield  {journal} {\bibinfo  {journal} {Biophys.
  Rev.}\ }\textbf {\bibinfo {volume} {9}},\ \bibinfo {pages} {699–728}
  (\bibinfo {year} {2017})}\BibitemShut {NoStop}%
\bibitem [{\citenamefont {B\"urgi}\ \emph {et~al.}(2002)\citenamefont
  {B\"urgi}, \citenamefont {Kollman},\ and\ \citenamefont {van
  Gunsteren}}]{Burgi2002}%
  \BibitemOpen
  \bibfield  {author} {\bibinfo {author} {\bibfnamefont {R.}~\bibnamefont
  {B\"urgi}}, \bibinfo {author} {\bibfnamefont {P.~A.}\ \bibnamefont
  {Kollman}}, \ and\ \bibinfo {author} {\bibfnamefont {W.~F.}\ \bibnamefont
  {van Gunsteren}},\ }\href@noop {} {\bibfield  {journal} {\bibinfo  {journal}
  {Proteins}\ }\textbf {\bibinfo {volume} {47}},\ \bibinfo {pages} {469}
  (\bibinfo {year} {2002})}\BibitemShut {NoStop}%
\bibitem [{\citenamefont {Mongan}\ \emph {et~al.}(2004)\citenamefont {Mongan},
  \citenamefont {Case},\ and\ \citenamefont {McCammon}}]{Mongan2004}%
  \BibitemOpen
  \bibfield  {author} {\bibinfo {author} {\bibfnamefont {J.}~\bibnamefont
  {Mongan}}, \bibinfo {author} {\bibfnamefont {D.~A.}\ \bibnamefont {Case}}, \
  and\ \bibinfo {author} {\bibfnamefont {J.~A.}\ \bibnamefont {McCammon}},\
  }\href@noop {} {\bibfield  {journal} {\bibinfo  {journal} {J. Comp. Chem.}\
  }\textbf {\bibinfo {volume} {25}},\ \bibinfo {pages} {2038} (\bibinfo {year}
  {2004})}\BibitemShut {NoStop}%
\bibitem [{\citenamefont {Williams}\ \emph {et~al.}(2010)\citenamefont
  {Williams}, \citenamefont {de~Oliveira},\ and\ \citenamefont
  {McCammon}}]{Williams2010}%
  \BibitemOpen
  \bibfield  {author} {\bibinfo {author} {\bibfnamefont {S.~L.}\ \bibnamefont
  {Williams}}, \bibinfo {author} {\bibfnamefont {C.~A.~F.}\ \bibnamefont
  {de~Oliveira}}, \ and\ \bibinfo {author} {\bibfnamefont {J.~A.}\ \bibnamefont
  {McCammon}},\ }\href@noop {} {\bibfield  {journal} {\bibinfo  {journal} {J.
  Chem. Theory Comput.}\ }\textbf {\bibinfo {volume} {6}},\ \bibinfo {pages}
  {560} (\bibinfo {year} {2010})}\BibitemShut {NoStop}%
\bibitem [{\citenamefont {Baptista}\ \emph {et~al.}(2002)\citenamefont
  {Baptista}, \citenamefont {Teixeira},\ and\ \citenamefont
  {Soares}}]{Baptista2002}%
  \BibitemOpen
  \bibfield  {author} {\bibinfo {author} {\bibfnamefont {A.~M.}\ \bibnamefont
  {Baptista}}, \bibinfo {author} {\bibfnamefont {V.~H.}\ \bibnamefont
  {Teixeira}}, \ and\ \bibinfo {author} {\bibfnamefont {C.~M.}\ \bibnamefont
  {Soares}},\ }\href@noop {} {\bibfield  {journal} {\bibinfo  {journal} {J.
  Chem. Phys.}\ }\textbf {\bibinfo {volume} {117}},\ \bibinfo {pages} {4184}
  (\bibinfo {year} {2002})}\BibitemShut {NoStop}%
\bibitem [{\citenamefont {Meng}\ and\ \citenamefont
  {Roitberg}(2010)}]{Meng2010}%
  \BibitemOpen
  \bibfield  {author} {\bibinfo {author} {\bibfnamefont {Y.}~\bibnamefont
  {Meng}}\ and\ \bibinfo {author} {\bibfnamefont {A.~E.}\ \bibnamefont
  {Roitberg}},\ }\href@noop {} {\bibfield  {journal} {\bibinfo  {journal} {J.
  Chem. Theory Comput.}\ }\textbf {\bibinfo {volume} {6}},\ \bibinfo {pages}
  {1401} (\bibinfo {year} {2010})}\BibitemShut {NoStop}%
\bibitem [{\citenamefont {Sabri~Dashti}\ \emph {et~al.}(2012)\citenamefont
  {Sabri~Dashti}, \citenamefont {Meng},\ and\ \citenamefont
  {Roitberg}}]{Dashti2012}%
  \BibitemOpen
  \bibfield  {author} {\bibinfo {author} {\bibfnamefont {D.}~\bibnamefont
  {Sabri~Dashti}}, \bibinfo {author} {\bibfnamefont {Y.}~\bibnamefont {Meng}},
  \ and\ \bibinfo {author} {\bibfnamefont {A.~E.}\ \bibnamefont {Roitberg}},\
  }\href@noop {} {\bibfield  {journal} {\bibinfo  {journal} {J. Phys. Chem. B}\
  }\textbf {\bibinfo {volume} {116}},\ \bibinfo {pages} {8805} (\bibinfo {year}
  {2012})}\BibitemShut {NoStop}%
\bibitem [{\citenamefont {Swails}\ \emph {et~al.}(2014)\citenamefont {Swails},
  \citenamefont {York},\ and\ \citenamefont {Roitberg}}]{Swails2014}%
  \BibitemOpen
  \bibfield  {author} {\bibinfo {author} {\bibfnamefont {J.~M.}\ \bibnamefont
  {Swails}}, \bibinfo {author} {\bibfnamefont {D.~M.}\ \bibnamefont {York}}, \
  and\ \bibinfo {author} {\bibfnamefont {A.~E.}\ \bibnamefont {Roitberg}},\
  }\href@noop {} {\bibfield  {journal} {\bibinfo  {journal} {J. Chem. Theory
  Comp.}\ }\textbf {\bibinfo {volume} {10}},\ \bibinfo {pages} {1341} (\bibinfo
  {year} {2014})}\BibitemShut {NoStop}%
\bibitem [{\citenamefont {Chen}\ and\ \citenamefont {Roux}(2015)}]{Chen2015}%
  \BibitemOpen
  \bibfield  {author} {\bibinfo {author} {\bibfnamefont {Y.}~\bibnamefont
  {Chen}}\ and\ \bibinfo {author} {\bibfnamefont {B.}~\bibnamefont {Roux}},\
  }\href@noop {} {\bibfield  {journal} {\bibinfo  {journal} {J. Chem. Theory
  Comput.}\ }\textbf {\bibinfo {volume} {11}},\ \bibinfo {pages} {3919}
  (\bibinfo {year} {2015})}\BibitemShut {NoStop}%
\bibitem [{\citenamefont {Stern}(2007)}]{Stern2007}%
  \BibitemOpen
  \bibfield  {author} {\bibinfo {author} {\bibfnamefont {H.~A.}\ \bibnamefont
  {Stern}},\ }\href@noop {} {\bibfield  {journal} {\bibinfo  {journal} {J.
  Chem. Phys.}\ }\textbf {\bibinfo {volume} {126}},\ \bibinfo {pages} {164112}
  (\bibinfo {year} {2007})}\BibitemShut {NoStop}%
\bibitem [{\citenamefont {Radak}\ \emph {et~al.}(2017)\citenamefont {Radak},
  \citenamefont {Chipot}, \citenamefont {Suh}, \citenamefont {Jo},
  \citenamefont {Jiang}, \citenamefont {Phillips}, \citenamefont {Schulten},\
  and\ \citenamefont {Roux}}]{Radak2017}%
  \BibitemOpen
  \bibfield  {author} {\bibinfo {author} {\bibfnamefont {B.~K.}\ \bibnamefont
  {Radak}}, \bibinfo {author} {\bibfnamefont {C.}~\bibnamefont {Chipot}},
  \bibinfo {author} {\bibfnamefont {D.}~\bibnamefont {Suh}}, \bibinfo {author}
  {\bibfnamefont {S.}~\bibnamefont {Jo}}, \bibinfo {author} {\bibfnamefont
  {W.}~\bibnamefont {Jiang}}, \bibinfo {author} {\bibfnamefont {J.~C.}\
  \bibnamefont {Phillips}}, \bibinfo {author} {\bibfnamefont {K.}~\bibnamefont
  {Schulten}}, \ and\ \bibinfo {author} {\bibfnamefont {B.}~\bibnamefont
  {Roux}},\ }\href@noop {} {\bibfield  {journal} {\bibinfo  {journal} {J. Chem.
  Theory Comput.}\ }\textbf {\bibinfo {volume} {13}},\ \bibinfo {pages} {5933}
  (\bibinfo {year} {2017})}\BibitemShut {NoStop}%
\bibitem [{\citenamefont {Sch{\"{u}}tte}\ \emph {et~al.}(1999)\citenamefont
  {Sch{\"{u}}tte}, \citenamefont {Huisinga},\ and\ \citenamefont
  {Deuflhard}}]{Schuette1999b}%
  \BibitemOpen
  \bibfield  {author} {\bibinfo {author} {\bibfnamefont {C.}~\bibnamefont
  {Sch{\"{u}}tte}}, \bibinfo {author} {\bibfnamefont {W.}~\bibnamefont
  {Huisinga}}, \ and\ \bibinfo {author} {\bibfnamefont {P.}~\bibnamefont
  {Deuflhard}},\ }\href@noop {} {\bibfield  {journal} {\bibinfo  {journal}
  {Z.I.B. Report}\ }\textbf {\bibinfo {volume} {36}},\ \bibinfo {pages} {191}
  (\bibinfo {year} {1999})}\BibitemShut {NoStop}%
\bibitem [{\citenamefont {Deuflhard}\ \emph {et~al.}(2000)\citenamefont
  {Deuflhard}, \citenamefont {Huisinga}, \citenamefont {Fischer},\ and\
  \citenamefont {Sch{\"{u}}tte}}]{Deuflhard2000}%
  \BibitemOpen
  \bibfield  {author} {\bibinfo {author} {\bibfnamefont {P.}~\bibnamefont
  {Deuflhard}}, \bibinfo {author} {\bibfnamefont {W.}~\bibnamefont {Huisinga}},
  \bibinfo {author} {\bibfnamefont {A.}~\bibnamefont {Fischer}}, \ and\
  \bibinfo {author} {\bibfnamefont {C.}~\bibnamefont {Sch{\"{u}}tte}},\ }\href
  {http://linkinghub.elsevier.com/retrieve/pii/S0024379500000951} {\bibfield
  {journal} {\bibinfo  {journal} {Linear Algebra Appl.}\ }\textbf {\bibinfo
  {volume} {315}},\ \bibinfo {pages} {39} (\bibinfo {year} {2000})}\BibitemShut
  {NoStop}%
\bibitem [{\citenamefont {Chodera}\ \emph {et~al.}(2007)\citenamefont
  {Chodera}, \citenamefont {Singhal}, \citenamefont {Pande}, \citenamefont
  {Dill},\ and\ \citenamefont {Swope}}]{Chodera2007}%
  \BibitemOpen
  \bibfield  {author} {\bibinfo {author} {\bibfnamefont {J.~D.}\ \bibnamefont
  {Chodera}}, \bibinfo {author} {\bibfnamefont {N.}~\bibnamefont {Singhal}},
  \bibinfo {author} {\bibfnamefont {V.~S.}\ \bibnamefont {Pande}}, \bibinfo
  {author} {\bibfnamefont {K.~A.}\ \bibnamefont {Dill}}, \ and\ \bibinfo
  {author} {\bibfnamefont {W.~C.}\ \bibnamefont {Swope}},\ }\href@noop {}
  {\bibfield  {journal} {\bibinfo  {journal} {J. Chem. Phys.}\ }\textbf
  {\bibinfo {volume} {126}},\ \bibinfo {pages} {155101} (\bibinfo {year}
  {2007})}\BibitemShut {NoStop}%
\bibitem [{\citenamefont {Buchete}\ and\ \citenamefont
  {Hummer}(2008)}]{Buchete2008}%
  \BibitemOpen
  \bibfield  {author} {\bibinfo {author} {\bibfnamefont {N.-V.}\ \bibnamefont
  {Buchete}}\ and\ \bibinfo {author} {\bibfnamefont {G.}~\bibnamefont
  {Hummer}},\ }\href@noop {} {\bibfield  {journal} {\bibinfo  {journal} {J.
  Phys. Chem. B}\ }\textbf {\bibinfo {volume} {112}},\ \bibinfo {pages} {6057}
  (\bibinfo {year} {2008})}\BibitemShut {NoStop}%
\bibitem [{\citenamefont {Prinz}\ \emph {et~al.}(2011)\citenamefont {Prinz},
  \citenamefont {Wu}, \citenamefont {Sarich}, \citenamefont {Keller},
  \citenamefont {Senne}, \citenamefont {Held}, \citenamefont {Chodera},
  \citenamefont {Sch{\"{u}}tte},\ and\ \citenamefont {No{\'{e}}}}]{Prinz2011}%
  \BibitemOpen
  \bibfield  {author} {\bibinfo {author} {\bibfnamefont {J.-H.}\ \bibnamefont
  {Prinz}}, \bibinfo {author} {\bibfnamefont {H.}~\bibnamefont {Wu}}, \bibinfo
  {author} {\bibfnamefont {M.}~\bibnamefont {Sarich}}, \bibinfo {author}
  {\bibfnamefont {B.}~\bibnamefont {Keller}}, \bibinfo {author} {\bibfnamefont
  {M.}~\bibnamefont {Senne}}, \bibinfo {author} {\bibfnamefont
  {M.}~\bibnamefont {Held}}, \bibinfo {author} {\bibfnamefont {J.~D.}\
  \bibnamefont {Chodera}}, \bibinfo {author} {\bibfnamefont {C.}~\bibnamefont
  {Sch{\"{u}}tte}}, \ and\ \bibinfo {author} {\bibfnamefont {F.}~\bibnamefont
  {No{\'{e}}}},\ }\href@noop {} {\bibfield  {journal} {\bibinfo  {journal} {J.
  Chem. Phys.}\ }\textbf {\bibinfo {volume} {134}},\ \bibinfo {pages} {174105}
  (\bibinfo {year} {2011})}\BibitemShut {NoStop}%
\bibitem [{\citenamefont {Keller}\ \emph {et~al.}(2011)\citenamefont {Keller},
  \citenamefont {H{\"{u}}nenberger},\ and\ \citenamefont {van
  Gunsteren}}]{Keller2011}%
  \BibitemOpen
  \bibfield  {author} {\bibinfo {author} {\bibfnamefont {B.}~\bibnamefont
  {Keller}}, \bibinfo {author} {\bibfnamefont {P.}~\bibnamefont
  {H{\"{u}}nenberger}}, \ and\ \bibinfo {author} {\bibfnamefont {W.~F.}\
  \bibnamefont {van Gunsteren}},\ }\href@noop {} {\bibfield  {journal}
  {\bibinfo  {journal} {J. Chem. Theory Comput.}\ }\textbf {\bibinfo {volume}
  {7}},\ \bibinfo {pages} {1032} (\bibinfo {year} {2011})}\BibitemShut
  {NoStop}%
\bibitem [{\citenamefont {Keller}\ \emph {et~al.}(2019)\citenamefont {Keller},
  \citenamefont {Aleksic},\ and\ \citenamefont {Donati}}]{keller2019}%
  \BibitemOpen
  \bibfield  {author} {\bibinfo {author} {\bibfnamefont {B.~G.}\ \bibnamefont
  {Keller}}, \bibinfo {author} {\bibfnamefont {S.}~\bibnamefont {Aleksic}}, \
  and\ \bibinfo {author} {\bibfnamefont {L.}~\bibnamefont {Donati}},\ }in\
  \href@noop {} {\emph {\bibinfo {booktitle} {Biomolecular Simulations in
  Structure-based Drug Discovery}}},\ \bibinfo {editor} {edited by\ \bibinfo
  {editor} {\bibfnamefont {F.~L.}\ \bibnamefont {Gervasio}}}\ (\bibinfo
  {publisher} {Wiley-Interscience},\ \bibinfo {address} {Weinheim},\ \bibinfo
  {year} {2019})\ p.~\bibinfo {pages} {67}\BibitemShut {NoStop}%
\bibitem [{\citenamefont {Deuflhard}\ and\ \citenamefont
  {Weber}(2004)}]{Deuflhard2004}%
  \BibitemOpen
  \bibfield  {author} {\bibinfo {author} {\bibfnamefont {P.}~\bibnamefont
  {Deuflhard}}\ and\ \bibinfo {author} {\bibfnamefont {M.}~\bibnamefont
  {Weber}},\ }\href@noop {} {\bibfield  {journal} {\bibinfo  {journal} {Linear
  Algebra Appl.}\ }\textbf {\bibinfo {volume} {398}},\ \bibinfo {pages} {161}
  (\bibinfo {year} {2004})}\BibitemShut {NoStop}%
\bibitem [{\citenamefont {Kube}\ and\ \citenamefont {Weber}(2007)}]{Kube2007}%
  \BibitemOpen
  \bibfield  {author} {\bibinfo {author} {\bibfnamefont {S.}~\bibnamefont
  {Kube}}\ and\ \bibinfo {author} {\bibfnamefont {M.}~\bibnamefont {Weber}},\
  }\href@noop {} {\bibfield  {journal} {\bibinfo  {journal} {J. Chem. Phys.}\
  }\textbf {\bibinfo {volume} {126}},\ \bibinfo {pages} {024103} (\bibinfo
  {year} {2007})}\BibitemShut {NoStop}%
\bibitem [{\citenamefont {R\"oblitz}\ and\ \citenamefont
  {Weber}(2013)}]{Roblitz2013}%
  \BibitemOpen
  \bibfield  {author} {\bibinfo {author} {\bibfnamefont {S.}~\bibnamefont
  {R\"oblitz}}\ and\ \bibinfo {author} {\bibfnamefont {M.}~\bibnamefont
  {Weber}},\ }\href@noop {} {\bibfield  {journal} {\bibinfo  {journal} {Adv.
  Data Anal. Classif.}\ }\textbf {\bibinfo {volume} {7}} (\bibinfo {year}
  {2013})}\BibitemShut {NoStop}%
\bibitem [{\citenamefont {Weber}(2018)}]{Weber2018}%
  \BibitemOpen
  \bibfield  {author} {\bibinfo {author} {\bibfnamefont {M.}~\bibnamefont
  {Weber}},\ }\href@noop {} {\bibfield  {journal} {\bibinfo  {journal}
  {Computation}\ }\textbf {\bibinfo {volume} {6}} (\bibinfo {year}
  {2018})}\BibitemShut {NoStop}%
\bibitem [{\citenamefont {Erlekam}\ \emph {et~al.}(2019)\citenamefont
  {Erlekam}, \citenamefont {Igde}, \citenamefont {Röblitz}, \citenamefont
  {Hartmann},\ and\ \citenamefont {Weber}}]{Erlekam2019}%
  \BibitemOpen
  \bibfield  {author} {\bibinfo {author} {\bibfnamefont {F.}~\bibnamefont
  {Erlekam}}, \bibinfo {author} {\bibfnamefont {S.}~\bibnamefont {Igde}},
  \bibinfo {author} {\bibfnamefont {S.}~\bibnamefont {Röblitz}}, \bibinfo
  {author} {\bibfnamefont {L.}~\bibnamefont {Hartmann}}, \ and\ \bibinfo
  {author} {\bibfnamefont {M.}~\bibnamefont {Weber}},\ }\href@noop {}
  {\bibfield  {journal} {\bibinfo  {journal} {Computation}\ }\textbf {\bibinfo
  {volume} {7}} (\bibinfo {year} {2019})}\BibitemShut {NoStop}%
\bibitem [{\citenamefont {Bicout}\ and\ \citenamefont
  {Szabo}(1998)}]{Bicout1998}%
  \BibitemOpen
  \bibfield  {author} {\bibinfo {author} {\bibfnamefont {D.~J.}\ \bibnamefont
  {Bicout}}\ and\ \bibinfo {author} {\bibfnamefont {A.}~\bibnamefont {Szabo}},\
  }\href@noop {} {\bibfield  {journal} {\bibinfo  {journal} {J. Chem. Phys.}\
  }\textbf {\bibinfo {volume} {109}},\ \bibinfo {pages} {10.1063/1.476800}
  (\bibinfo {year} {1998})}\BibitemShut {NoStop}%
\bibitem [{\citenamefont {Lie}\ \emph {et~al.}(2013)\citenamefont {Lie},
  \citenamefont {Fackeldey},\ and\ \citenamefont {Weber}}]{Lie2013}%
  \BibitemOpen
  \bibfield  {author} {\bibinfo {author} {\bibfnamefont {H.~C.}\ \bibnamefont
  {Lie}}, \bibinfo {author} {\bibfnamefont {K.}~\bibnamefont {Fackeldey}}, \
  and\ \bibinfo {author} {\bibfnamefont {M.}~\bibnamefont {Weber}},\
  }\href@noop {} {\bibfield  {journal} {\bibinfo  {journal} {SIAM. J. Matrix
  Anal. Appl.}\ }\textbf {\bibinfo {volume} {34}},\ \bibinfo {pages}
  {738–756} (\bibinfo {year} {2013})}\BibitemShut {NoStop}%
\bibitem [{\citenamefont {Donati}\ \emph {et~al.}(2018)\citenamefont {Donati},
  \citenamefont {Heida}, \citenamefont {Keller},\ and\ \citenamefont
  {Weber}}]{Donati2018b}%
  \BibitemOpen
  \bibfield  {author} {\bibinfo {author} {\bibfnamefont {L.}~\bibnamefont
  {Donati}}, \bibinfo {author} {\bibfnamefont {M.}~\bibnamefont {Heida}},
  \bibinfo {author} {\bibfnamefont {B.~G.}\ \bibnamefont {Keller}}, \ and\
  \bibinfo {author} {\bibfnamefont {M.}~\bibnamefont {Weber}},\ }\href@noop {}
  {\bibfield  {journal} {\bibinfo  {journal} {J. Phys. Condens. Matter}\
  }\textbf {\bibinfo {volume} {30}},\ \bibinfo {pages} {425201} (\bibinfo
  {year} {2018})}\BibitemShut {NoStop}%
\bibitem [{\citenamefont {Dixit}\ \emph {et~al.}(2015)\citenamefont {Dixit},
  \citenamefont {Jain}, \citenamefont {Stock},\ and\ \citenamefont
  {Dill}}]{Dixit2015}%
  \BibitemOpen
  \bibfield  {author} {\bibinfo {author} {\bibfnamefont {P.~D.}\ \bibnamefont
  {Dixit}}, \bibinfo {author} {\bibfnamefont {A.}~\bibnamefont {Jain}},
  \bibinfo {author} {\bibfnamefont {G.}~\bibnamefont {Stock}}, \ and\ \bibinfo
  {author} {\bibfnamefont {K.~A.}\ \bibnamefont {Dill}},\ }\href@noop {}
  {\bibfield  {journal} {\bibinfo  {journal} {J. Chem. Theory Comput.}\
  }\textbf {\bibinfo {volume} {11}},\ \bibinfo {pages} {5464–5472} (\bibinfo
  {year} {2015})}\BibitemShut {NoStop}%
\bibitem [{\citenamefont {Stock}\ \emph {et~al.}(2008)\citenamefont {Stock},
  \citenamefont {Ghosh},\ and\ \citenamefont {Dill}}]{Stock2008}%
  \BibitemOpen
  \bibfield  {author} {\bibinfo {author} {\bibfnamefont {G.}~\bibnamefont
  {Stock}}, \bibinfo {author} {\bibfnamefont {K.}~\bibnamefont {Ghosh}}, \ and\
  \bibinfo {author} {\bibfnamefont {K.~A.}\ \bibnamefont {Dill}},\ }\href@noop
  {} {\bibfield  {journal} {\bibinfo  {journal} {J. Chem. Phys.}\ }\textbf
  {\bibinfo {volume} {128}},\ \bibinfo {pages} {194102} (\bibinfo {year}
  {2008})}\BibitemShut {NoStop}%
\bibitem [{\citenamefont {Otten}\ and\ \citenamefont
  {Stock}(2010)}]{Otten2010}%
  \BibitemOpen
  \bibfield  {author} {\bibinfo {author} {\bibfnamefont {M.}~\bibnamefont
  {Otten}}\ and\ \bibinfo {author} {\bibfnamefont {G.}~\bibnamefont {Stock}},\
  }\href@noop {} {\bibfield  {journal} {\bibinfo  {journal} {J. Chem. Phys.}\
  }\textbf {\bibinfo {volume} {133}},\ \bibinfo {pages} {034119} (\bibinfo
  {year} {2010})}\BibitemShut {NoStop}%
\bibitem [{\citenamefont {Rosta}\ and\ \citenamefont
  {Hummer}(2015)}]{Rosta2015}%
  \BibitemOpen
  \bibfield  {author} {\bibinfo {author} {\bibfnamefont {E.}~\bibnamefont
  {Rosta}}\ and\ \bibinfo {author} {\bibfnamefont {G.}~\bibnamefont {Hummer}},\
  }\href@noop {} {\bibfield  {journal} {\bibinfo  {journal} {J. Chem. Theory
  Comput.}\ }\textbf {\bibinfo {volume} {11}},\ \bibinfo {pages} {276–285}
  (\bibinfo {year} {2015})}\BibitemShut {NoStop}%
\bibitem [{\citenamefont {Donati}\ \emph {et~al.}(2021)\citenamefont {Donati},
  \citenamefont {Weber},\ and\ \citenamefont {Keller}}]{Donati2021}%
  \BibitemOpen
  \bibfield  {author} {\bibinfo {author} {\bibfnamefont {L.}~\bibnamefont
  {Donati}}, \bibinfo {author} {\bibfnamefont {M.}~\bibnamefont {Weber}}, \
  and\ \bibinfo {author} {\bibfnamefont {B.~G.}\ \bibnamefont {Keller}},\
  }\href@noop {} {\bibfield  {journal} {\bibinfo  {journal} {J. Phys. Condens.
  Matter}\ }\textbf {\bibinfo {volume} {33}},\ \bibinfo {pages} {115902}
  (\bibinfo {year} {2021})}\BibitemShut {NoStop}%
\bibitem [{\citenamefont {Ray}\ \emph {et~al.}(2020)\citenamefont {Ray},
  \citenamefont {Sunkara}, \citenamefont {Schütte},\ and\ \citenamefont
  {Weber}}]{Ray2020}%
  \BibitemOpen
  \bibfield  {author} {\bibinfo {author} {\bibfnamefont {S.}~\bibnamefont
  {Ray}}, \bibinfo {author} {\bibfnamefont {V.}~\bibnamefont {Sunkara}},
  \bibinfo {author} {\bibfnamefont {C.}~\bibnamefont {Schütte}}, \ and\
  \bibinfo {author} {\bibfnamefont {M.}~\bibnamefont {Weber}},\ }\href@noop {}
  {\bibfield  {journal} {\bibinfo  {journal} {Mol. Simul.}\ }\textbf {\bibinfo
  {volume} {46}},\ \bibinfo {pages} {1443} (\bibinfo {year}
  {2020})}\BibitemShut {NoStop}%
\bibitem [{\citenamefont {Spahn}\ \emph {et~al.}(2017)\citenamefont {Spahn},
  \citenamefont {Vecchio}, \citenamefont {Labuz}, \citenamefont
  {Rodriguez-Gaztelumendi}, \citenamefont {Massaly}, \citenamefont {Temp},
  \citenamefont {Durmaz}, \citenamefont {Sabri}, \citenamefont {Reidelbach},
  \citenamefont {Machelska}, \citenamefont {Weber},\ and\ \citenamefont
  {Stein}}]{Spahn2017}%
  \BibitemOpen
  \bibfield  {author} {\bibinfo {author} {\bibfnamefont {V.}~\bibnamefont
  {Spahn}}, \bibinfo {author} {\bibfnamefont {G.~D.}\ \bibnamefont {Vecchio}},
  \bibinfo {author} {\bibfnamefont {D.}~\bibnamefont {Labuz}}, \bibinfo
  {author} {\bibfnamefont {A.}~\bibnamefont {Rodriguez-Gaztelumendi}}, \bibinfo
  {author} {\bibfnamefont {N.}~\bibnamefont {Massaly}}, \bibinfo {author}
  {\bibfnamefont {J.}~\bibnamefont {Temp}}, \bibinfo {author} {\bibfnamefont
  {V.}~\bibnamefont {Durmaz}}, \bibinfo {author} {\bibfnamefont
  {P.}~\bibnamefont {Sabri}}, \bibinfo {author} {\bibfnamefont
  {M.}~\bibnamefont {Reidelbach}}, \bibinfo {author} {\bibfnamefont
  {H.}~\bibnamefont {Machelska}}, \bibinfo {author} {\bibfnamefont
  {M.}~\bibnamefont {Weber}}, \ and\ \bibinfo {author} {\bibfnamefont
  {C.}~\bibnamefont {Stein}},\ }\href@noop {} {\bibfield  {journal} {\bibinfo
  {journal} {Science}\ }\textbf {\bibinfo {volume} {355}},\ \bibinfo {pages}
  {966} (\bibinfo {year} {2017})}\BibitemShut {NoStop}%
\bibitem [{\citenamefont {Lynch}\ and\ \citenamefont
  {Pettitt}(1997)}]{Lynch1997}%
  \BibitemOpen
  \bibfield  {author} {\bibinfo {author} {\bibfnamefont {G.~C.}\ \bibnamefont
  {Lynch}}\ and\ \bibinfo {author} {\bibfnamefont {B.~M.}\ \bibnamefont
  {Pettitt}},\ }\href@noop {} {\bibfield  {journal} {\bibinfo  {journal} {J.
  Chem. Phys.}\ }\textbf {\bibinfo {volume} {107}},\ \bibinfo {pages} {8594}
  (\bibinfo {year} {1997})}\BibitemShut {NoStop}%
\bibitem [{\citenamefont {N\"ske}\ \emph {et~al.}(2014)\citenamefont {N\"ske},
  \citenamefont {Keller}, \citenamefont {Pérez-Hernández}, \citenamefont
  {Mey},\ and\ \citenamefont {Noé}}]{Nuske2014}%
  \BibitemOpen
  \bibfield  {author} {\bibinfo {author} {\bibfnamefont {F.}~\bibnamefont
  {N\"ske}}, \bibinfo {author} {\bibfnamefont {B.~G.}\ \bibnamefont {Keller}},
  \bibinfo {author} {\bibfnamefont {G.}~\bibnamefont {Pérez-Hernández}},
  \bibinfo {author} {\bibfnamefont {A.~S. J.~S.}\ \bibnamefont {Mey}}, \ and\
  \bibinfo {author} {\bibfnamefont {F.}~\bibnamefont {Noé}},\ }\href {\doibase
  10.1021/ct4009156} {\bibfield  {journal} {\bibinfo  {journal} {J. Chem.
  Theory Comput.}\ }\textbf {\bibinfo {volume} {10}},\ \bibinfo {pages} {1739}
  (\bibinfo {year} {2014})},\ \bibinfo {note} {pMID: 26580382}\BibitemShut
  {NoStop}%
\bibitem [{\citenamefont {Kieninger}\ and\ \citenamefont
  {Keller}(2021)}]{Kieninger2021}%
  \BibitemOpen
  \bibfield  {author} {\bibinfo {author} {\bibfnamefont {S.}~\bibnamefont
  {Kieninger}}\ and\ \bibinfo {author} {\bibfnamefont {B.~G.}\ \bibnamefont
  {Keller}},\ }\href@noop {} {\bibfield  {journal} {\bibinfo  {journal} {J.
  Chem. Phys.}\ }\textbf {\bibinfo {volume} {154}},\ \bibinfo {pages} {094102}
  (\bibinfo {year} {2021})}\BibitemShut {NoStop}%
\bibitem [{\citenamefont {Mori}(1965)}]{Mori1965}%
  \BibitemOpen
  \bibfield  {author} {\bibinfo {author} {\bibfnamefont {H.}~\bibnamefont
  {Mori}},\ }\href@noop {} {\bibfield  {journal} {\bibinfo  {journal} {Prog.
  Theor. Phys}\ }\textbf {\bibinfo {volume} {33}},\ \bibinfo {pages} {423}
  (\bibinfo {year} {1965})}\BibitemShut {NoStop}%
\bibitem [{\citenamefont {Zwanzig}(1973)}]{Zwanzig1973}%
  \BibitemOpen
  \bibfield  {author} {\bibinfo {author} {\bibfnamefont {R.}~\bibnamefont
  {Zwanzig}},\ }\href@noop {} {\bibfield  {journal} {\bibinfo  {journal} {J.
  Stat. Phys.}\ }\textbf {\bibinfo {volume} {9}},\ \bibinfo {pages} {215–220}
  (\bibinfo {year} {1973})}\BibitemShut {NoStop}%
\bibitem [{\citenamefont {Van~Kampen}(1985)}]{Kampen1985}%
  \BibitemOpen
  \bibfield  {author} {\bibinfo {author} {\bibfnamefont {N.~G.}\ \bibnamefont
  {Van~Kampen}},\ }\href@noop {} {\bibfield  {journal} {\bibinfo  {journal}
  {Phys. Rep.}\ }\textbf {\bibinfo {volume} {124}},\ \bibinfo {pages} {69}
  (\bibinfo {year} {1985})}\BibitemShut {NoStop}%
\bibitem [{\citenamefont {Straub}\ \emph {et~al.}(1987)\citenamefont {Straub},
  \citenamefont {Borkovec},\ and\ \citenamefont {Berne}}]{straub1987}%
  \BibitemOpen
  \bibfield  {author} {\bibinfo {author} {\bibfnamefont {J.~E.}\ \bibnamefont
  {Straub}}, \bibinfo {author} {\bibfnamefont {M.}~\bibnamefont {Borkovec}}, \
  and\ \bibinfo {author} {\bibfnamefont {B.~J.}\ \bibnamefont {Berne}},\
  }\href@noop {} {\bibfield  {journal} {\bibinfo  {journal} {J. Chem. Phys.}\
  }\textbf {\bibinfo {volume} {91}},\ \bibinfo {pages} {4995} (\bibinfo {year}
  {1987})}\BibitemShut {NoStop}%
\bibitem [{\citenamefont {Horenko}\ \emph {et~al.}(2007)\citenamefont
  {Horenko}, \citenamefont {Hartmann}, \citenamefont {Sch{\"u}tte},\ and\
  \citenamefont {Noe}}]{Horenko2007}%
  \BibitemOpen
  \bibfield  {author} {\bibinfo {author} {\bibfnamefont {I.}~\bibnamefont
  {Horenko}}, \bibinfo {author} {\bibfnamefont {C.}~\bibnamefont {Hartmann}},
  \bibinfo {author} {\bibfnamefont {C.}~\bibnamefont {Sch{\"u}tte}}, \ and\
  \bibinfo {author} {\bibfnamefont {F.}~\bibnamefont {Noe}},\ }\href@noop {}
  {\bibfield  {journal} {\bibinfo  {journal} {Phys. Rev. E}\ }\textbf {\bibinfo
  {volume} {76}},\ \bibinfo {pages} {016706} (\bibinfo {year}
  {2007})}\BibitemShut {NoStop}%
\bibitem [{\citenamefont {Darve}\ \emph {et~al.}(2009)\citenamefont {Darve},
  \citenamefont {Solomon},\ and\ \citenamefont {Kia}}]{darve2009}%
  \BibitemOpen
  \bibfield  {author} {\bibinfo {author} {\bibfnamefont {E.}~\bibnamefont
  {Darve}}, \bibinfo {author} {\bibfnamefont {J.}~\bibnamefont {Solomon}}, \
  and\ \bibinfo {author} {\bibfnamefont {A.}~\bibnamefont {Kia}},\ }\href@noop
  {} {\bibfield  {journal} {\bibinfo  {journal} {Proc. Natl. Acad. Sci.
  U.S.A.}\ }\textbf {\bibinfo {volume} {106}},\ \bibinfo {pages} {10884}
  (\bibinfo {year} {2009})}\BibitemShut {NoStop}%
\bibitem [{\citenamefont {Jung}\ \emph {et~al.}(2017)\citenamefont {Jung},
  \citenamefont {Hanke},\ and\ \citenamefont {Schmid}}]{jung2017}%
  \BibitemOpen
  \bibfield  {author} {\bibinfo {author} {\bibfnamefont {G.}~\bibnamefont
  {Jung}}, \bibinfo {author} {\bibfnamefont {M.}~\bibnamefont {Hanke}}, \ and\
  \bibinfo {author} {\bibfnamefont {F.}~\bibnamefont {Schmid}},\ }\href@noop {}
  {\bibfield  {journal} {\bibinfo  {journal} {J. Chem. Theory Comput.}\
  }\textbf {\bibinfo {volume} {13}},\ \bibinfo {pages} {2481} (\bibinfo {year}
  {2017})}\BibitemShut {NoStop}%
\bibitem [{\citenamefont {Daldrop}\ \emph {et~al.}(2018)\citenamefont
  {Daldrop}, \citenamefont {Kappler}, \citenamefont {Br{\"u}nig},\ and\
  \citenamefont {Netz}}]{Daldrop2018}%
  \BibitemOpen
  \bibfield  {author} {\bibinfo {author} {\bibfnamefont {J.~O.}\ \bibnamefont
  {Daldrop}}, \bibinfo {author} {\bibfnamefont {J.}~\bibnamefont {Kappler}},
  \bibinfo {author} {\bibfnamefont {F.~N.}\ \bibnamefont {Br{\"u}nig}}, \ and\
  \bibinfo {author} {\bibfnamefont {R.~R.}\ \bibnamefont {Netz}},\ }\href@noop
  {} {\bibfield  {journal} {\bibinfo  {journal} {Proc. Natl. Acad. Sci.
  U.S.A.}\ }\textbf {\bibinfo {volume} {115}},\ \bibinfo {pages} {5169}
  (\bibinfo {year} {2018})}\BibitemShut {NoStop}%
\bibitem [{\citenamefont {Lee}\ \emph {et~al.}(2019)\citenamefont {Lee},
  \citenamefont {Ahn},\ and\ \citenamefont {Darve}}]{Lee2019}%
  \BibitemOpen
  \bibfield  {author} {\bibinfo {author} {\bibfnamefont {H.~S.}\ \bibnamefont
  {Lee}}, \bibinfo {author} {\bibfnamefont {S.-H.}\ \bibnamefont {Ahn}}, \ and\
  \bibinfo {author} {\bibfnamefont {E.~F.}\ \bibnamefont {Darve}},\ }\href@noop
  {} {\bibfield  {journal} {\bibinfo  {journal} {J. Chem. Phys.}\ }\textbf
  {\bibinfo {volume} {150}},\ \bibinfo {pages} {174113} (\bibinfo {year}
  {2019})}\BibitemShut {NoStop}%
\bibitem [{\citenamefont {Ayaz}\ \emph {et~al.}(2021)\citenamefont {Ayaz},
  \citenamefont {Tepper}, \citenamefont {Brünig}, \citenamefont {Kappler},
  \citenamefont {Daldrop},\ and\ \citenamefont {Netz}}]{Ayaz2020}%
  \BibitemOpen
  \bibfield  {author} {\bibinfo {author} {\bibfnamefont {C.}~\bibnamefont
  {Ayaz}}, \bibinfo {author} {\bibfnamefont {L.}~\bibnamefont {Tepper}},
  \bibinfo {author} {\bibfnamefont {F.~N.}\ \bibnamefont {Brünig}}, \bibinfo
  {author} {\bibfnamefont {J.}~\bibnamefont {Kappler}}, \bibinfo {author}
  {\bibfnamefont {J.~O.}\ \bibnamefont {Daldrop}}, \ and\ \bibinfo {author}
  {\bibfnamefont {R.~R.}\ \bibnamefont {Netz}},\ }\href {\doibase
  10.1073/pnas.2023856118} {\bibfield  {journal} {\bibinfo  {journal} {Proc.
  Natl. Acad. Sci. U.S.A.}\ }\textbf {\bibinfo {volume} {118}},\ \bibinfo
  {pages} {e2023856118} (\bibinfo {year} {2021})}\BibitemShut {NoStop}%
\bibitem [{\citenamefont {Best}\ and\ \citenamefont {Hummer}(2011)}]{Best2011}%
  \BibitemOpen
  \bibfield  {author} {\bibinfo {author} {\bibfnamefont {R.~B.}\ \bibnamefont
  {Best}}\ and\ \bibinfo {author} {\bibfnamefont {G.}~\bibnamefont {Hummer}},\
  }\href@noop {} {\bibfield  {journal} {\bibinfo  {journal} {Phys. Chem. Chem.
  Phys.}\ }\textbf {\bibinfo {volume} {13}},\ \bibinfo {pages} {16902}
  (\bibinfo {year} {2011})}\BibitemShut {NoStop}%
\bibitem [{\citenamefont {Leimkuhler}\ and\ \citenamefont
  {Matthews}(2015)}]{Leimkuhler2015}%
  \BibitemOpen
  \bibfield  {author} {\bibinfo {author} {\bibfnamefont {B.}~\bibnamefont
  {Leimkuhler}}\ and\ \bibinfo {author} {\bibfnamefont {C.}~\bibnamefont
  {Matthews}},\ }\href@noop {} {\emph {\bibinfo {title} {Molecular Dynamics:
  With Deterministic and Stochastic Numerical Methods.}}}\ (\bibinfo
  {publisher} {Springer},\ \bibinfo {address} {Interdisciplinary Applied
  Mathematics; Vol. 39},\ \bibinfo {year} {2015})\BibitemShut {NoStop}%
\bibitem [{\citenamefont {Milstein}\ and\ \citenamefont
  {Tretyakov}(2004)}]{Milstein2004}%
  \BibitemOpen
  \bibfield  {author} {\bibinfo {author} {\bibfnamefont {G.}~\bibnamefont
  {Milstein}}\ and\ \bibinfo {author} {\bibfnamefont {M.}~\bibnamefont
  {Tretyakov}},\ }\href@noop {} {\emph {\bibinfo {title} {Stochastic Numerics
  for Mathematical Physics.}}}\ (\bibinfo  {publisher} {Springer, New York},\
  \bibinfo {address} {Interdisciplinary Applied Mathematics; Vol. 39},\
  \bibinfo {year} {2004})\BibitemShut {NoStop}%
\bibitem [{\citenamefont {MacQueen}(1967)}]{MacQueen1967}%
  \BibitemOpen
  \bibfield  {author} {\bibinfo {author} {\bibfnamefont {J.}~\bibnamefont
  {MacQueen}},\ }in\ \href@noop {} {\emph {\bibinfo {booktitle} {5th Berkeley
  Symp. Math. Statist. Probability}}}\ (\bibinfo {year} {1967})\ pp.\ \bibinfo
  {pages} {281--297}\BibitemShut {NoStop}%
\bibitem [{\citenamefont {Best}\ and\ \citenamefont {Hummer}(2010)}]{Best2010}%
  \BibitemOpen
  \bibfield  {author} {\bibinfo {author} {\bibfnamefont {R.~B.}\ \bibnamefont
  {Best}}\ and\ \bibinfo {author} {\bibfnamefont {G.}~\bibnamefont {Hummer}},\
  }\href@noop {} {\bibfield  {journal} {\bibinfo  {journal} {Proc. Natl. Acad.
  Sci. U.S.A.}\ }\textbf {\bibinfo {volume} {107}},\ \bibinfo {pages}
  {1088–1093} (\bibinfo {year} {2010})}\BibitemShut {NoStop}%
\bibitem [{\citenamefont {Lee}\ \emph {et~al.}(2016)\citenamefont {Lee},
  \citenamefont {Comer}, \citenamefont {Herndon}, \citenamefont {Leung},
  \citenamefont {Pavlova}, \citenamefont {Swift}, \citenamefont {Tung},
  \citenamefont {Rowley}, \citenamefont {Amaro}, \citenamefont {Chipot},
  \citenamefont {Wang},\ and\ \citenamefont {Gumbart}}]{Lee2016}%
  \BibitemOpen
  \bibfield  {author} {\bibinfo {author} {\bibfnamefont {C.~T.}\ \bibnamefont
  {Lee}}, \bibinfo {author} {\bibfnamefont {J.}~\bibnamefont {Comer}}, \bibinfo
  {author} {\bibfnamefont {C.}~\bibnamefont {Herndon}}, \bibinfo {author}
  {\bibfnamefont {N.}~\bibnamefont {Leung}}, \bibinfo {author} {\bibfnamefont
  {A.}~\bibnamefont {Pavlova}}, \bibinfo {author} {\bibfnamefont {R.~V.}\
  \bibnamefont {Swift}}, \bibinfo {author} {\bibfnamefont {C.}~\bibnamefont
  {Tung}}, \bibinfo {author} {\bibfnamefont {C.~N.}\ \bibnamefont {Rowley}},
  \bibinfo {author} {\bibfnamefont {R.~E.}\ \bibnamefont {Amaro}}, \bibinfo
  {author} {\bibfnamefont {C.}~\bibnamefont {Chipot}}, \bibinfo {author}
  {\bibfnamefont {Y.}~\bibnamefont {Wang}}, \ and\ \bibinfo {author}
  {\bibfnamefont {J.~C.}\ \bibnamefont {Gumbart}},\ }\href@noop {} {\bibfield
  {journal} {\bibinfo  {journal} {J. Chem. Inf. Model.}\ }\textbf {\bibinfo
  {volume} {56}},\ \bibinfo {pages} {721} (\bibinfo {year} {2016})}\BibitemShut
  {NoStop}%
\bibitem [{\citenamefont {Woolf}\ and\ \citenamefont {Roux}(1994)}]{Woolf1994}%
  \BibitemOpen
  \bibfield  {author} {\bibinfo {author} {\bibfnamefont {T.}~\bibnamefont
  {Woolf}}\ and\ \bibinfo {author} {\bibfnamefont {B.}~\bibnamefont {Roux}},\
  }\href@noop {} {\bibfield  {journal} {\bibinfo  {journal} {J. Am. Chem.
  Soc.}\ }\textbf {\bibinfo {volume} {116}},\ \bibinfo {pages} {5916} (\bibinfo
  {year} {1994})}\BibitemShut {NoStop}%
\bibitem [{\citenamefont {Hummer}(2005)}]{Hummer2005}%
  \BibitemOpen
  \bibfield  {author} {\bibinfo {author} {\bibfnamefont {G.}~\bibnamefont
  {Hummer}},\ }\href@noop {} {\bibfield  {journal} {\bibinfo  {journal} {New J.
  Phys.}\ }\textbf {\bibinfo {volume} {7}},\ \bibinfo {pages} {34–34}
  (\bibinfo {year} {2005})}\BibitemShut {NoStop}%
\bibitem [{\citenamefont {Hinczewski}\ \emph {et~al.}(2010)\citenamefont
  {Hinczewski}, \citenamefont {von Hansen}, \citenamefont {Dzubiella},\ and\
  \citenamefont {Netz}}]{Hinczewski2010}%
  \BibitemOpen
  \bibfield  {author} {\bibinfo {author} {\bibfnamefont {M.}~\bibnamefont
  {Hinczewski}}, \bibinfo {author} {\bibfnamefont {Y.}~\bibnamefont {von
  Hansen}}, \bibinfo {author} {\bibfnamefont {J.}~\bibnamefont {Dzubiella}}, \
  and\ \bibinfo {author} {\bibfnamefont {R.~R.}\ \bibnamefont {Netz}},\
  }\href@noop {} {\bibfield  {journal} {\bibinfo  {journal} {J. Chem. Phys.}\
  }\textbf {\bibinfo {volume} {132}},\ \bibinfo {pages} {245103} (\bibinfo
  {year} {2010})}\BibitemShut {NoStop}%
\bibitem [{\citenamefont {{Sedlmeier}}\ \emph {et~al.}(2011)\citenamefont
  {{Sedlmeier}}, \citenamefont {{von Hansen}}, \citenamefont {{Mengyu}},
  \citenamefont {{Horinek}},\ and\ \citenamefont {{Netz}}}]{Sedlmeier2011}%
  \BibitemOpen
  \bibfield  {author} {\bibinfo {author} {\bibfnamefont {F.}~\bibnamefont
  {{Sedlmeier}}}, \bibinfo {author} {\bibfnamefont {Y.}~\bibnamefont {{von
  Hansen}}}, \bibinfo {author} {\bibfnamefont {L.}~\bibnamefont {{Mengyu}}},
  \bibinfo {author} {\bibfnamefont {D.}~\bibnamefont {{Horinek}}}, \ and\
  \bibinfo {author} {\bibfnamefont {R.~R.}\ \bibnamefont {{Netz}}},\
  }\href@noop {} {\bibfield  {journal} {\bibinfo  {journal} {J. Stat. Phys.}\
  }\textbf {\bibinfo {volume} {145}},\ \bibinfo {pages} {240} (\bibinfo {year}
  {2011})}\BibitemShut {NoStop}%
\bibitem [{\citenamefont {Sicard}\ \emph {et~al.}(2021)\citenamefont {Sicard},
  \citenamefont {Koskin}, \citenamefont {Annibale},\ and\ \citenamefont
  {Rosta}}]{Sicard2021}%
  \BibitemOpen
  \bibfield  {author} {\bibinfo {author} {\bibfnamefont {F.}~\bibnamefont
  {Sicard}}, \bibinfo {author} {\bibfnamefont {V.}~\bibnamefont {Koskin}},
  \bibinfo {author} {\bibfnamefont {A.}~\bibnamefont {Annibale}}, \ and\
  \bibinfo {author} {\bibfnamefont {E.}~\bibnamefont {Rosta}},\ }\href
  {\doibase 10.1021/acs.jctc.0c01151} {\bibfield  {journal} {\bibinfo
  {journal} {J. Chem. Theory Comp.}\ }\textbf {\bibinfo {volume} {17}},\
  \bibinfo {pages} {2022} (\bibinfo {year} {2021})}\BibitemShut {NoStop}%
\bibitem [{\citenamefont {{Heida, Martin}}\ \emph {et~al.}(2021)\citenamefont
  {{Heida, Martin}}, \citenamefont {{Kantner, Markus}},\ and\ \citenamefont
  {{Stephan, Artur}}}]{Heida2021}%
  \BibitemOpen
  \bibfield  {author} {\bibinfo {author} {\bibnamefont {{Heida, Martin}}},
  \bibinfo {author} {\bibnamefont {{Kantner, Markus}}}, \ and\ \bibinfo
  {author} {\bibnamefont {{Stephan, Artur}}},\ }\href {\doibase
  10.1051/m2an/2021078} {\bibfield  {journal} {\bibinfo  {journal} {ESAIM:
  M2AN}\ }\textbf {\bibinfo {volume} {55}},\ \bibinfo {pages} {3017} (\bibinfo
  {year} {2021})}\BibitemShut {NoStop}%
\bibitem [{\citenamefont {Heida}(2018)}]{Heida2018}%
  \BibitemOpen
  \bibfield  {author} {\bibinfo {author} {\bibfnamefont {M.}~\bibnamefont
  {Heida}},\ }\href@noop {} {\bibfield  {journal} {\bibinfo  {journal} {Math.
  Models Methods Appl. Sci.}\ }\textbf {\bibinfo {volume} {28}},\ \bibinfo
  {pages} {2599–2635} (\bibinfo {year} {2018})}\BibitemShut {NoStop}%
\bibitem [{\citenamefont {Kong}\ and\ \citenamefont
  {Brooks~III}(1996)}]{Kong1996}%
  \BibitemOpen
  \bibfield  {author} {\bibinfo {author} {\bibfnamefont {X.}~\bibnamefont
  {Kong}}\ and\ \bibinfo {author} {\bibfnamefont {C.~L.}\ \bibnamefont
  {Brooks~III}},\ }\href@noop {} {\bibfield  {journal} {\bibinfo  {journal} {J.
  Chem. Phys.}\ }\textbf {\bibinfo {volume} {105}},\ \bibinfo {pages} {2414}
  (\bibinfo {year} {1996})}\BibitemShut {NoStop}%
\bibitem [{\citenamefont {Aho}\ \emph {et~al.}(2022)\citenamefont {Aho},
  \citenamefont {Buslaev}, \citenamefont {Jansen}, \citenamefont {Bauer},
  \citenamefont {Groenhof},\ and\ \citenamefont {Hess}}]{Aho2022}%
  \BibitemOpen
  \bibfield  {author} {\bibinfo {author} {\bibfnamefont {N.}~\bibnamefont
  {Aho}}, \bibinfo {author} {\bibfnamefont {P.}~\bibnamefont {Buslaev}},
  \bibinfo {author} {\bibfnamefont {A.}~\bibnamefont {Jansen}}, \bibinfo
  {author} {\bibfnamefont {P.}~\bibnamefont {Bauer}}, \bibinfo {author}
  {\bibfnamefont {G.}~\bibnamefont {Groenhof}}, \ and\ \bibinfo {author}
  {\bibfnamefont {B.}~\bibnamefont {Hess}},\ }\href {\doibase
  10.1021/acs.jctc.2c00516} {\bibfield  {journal} {\bibinfo  {journal} {J.
  Chem. Theory Comput.}\ }\textbf {\bibinfo {volume} {18}},\ \bibinfo {pages}
  {6148} (\bibinfo {year} {2022})}\BibitemShut {NoStop}%
\bibitem [{\citenamefont {Prabhulkar}\ \emph {et~al.}(2012)\citenamefont
  {Prabhulkar}, \citenamefont {Tian}, \citenamefont {Wang}, \citenamefont
  {Zhu},\ and\ \citenamefont {zhong Li}}]{Prabhulkar2012}%
  \BibitemOpen
  \bibfield  {author} {\bibinfo {author} {\bibfnamefont {S.}~\bibnamefont
  {Prabhulkar}}, \bibinfo {author} {\bibfnamefont {H.}~\bibnamefont {Tian}},
  \bibinfo {author} {\bibfnamefont {X.}~\bibnamefont {Wang}}, \bibinfo {author}
  {\bibfnamefont {J.}~\bibnamefont {Zhu}}, \ and\ \bibinfo {author}
  {\bibfnamefont {C.}~\bibnamefont {zhong Li}},\ }\href@noop {} {\bibfield
  {journal} {\bibinfo  {journal} {Antioxid. Redox Signal.}\ }\textbf {\bibinfo
  {volume} {17 12}},\ \bibinfo {pages} {1796} (\bibinfo {year}
  {2012})}\BibitemShut {NoStop}%
\bibitem [{\citenamefont {Shao}\ \emph {et~al.}(2022)\citenamefont {Shao},
  \citenamefont {Zhang},\ and\ \citenamefont {Xu}}]{Shao2022}%
  \BibitemOpen
  \bibfield  {author} {\bibinfo {author} {\bibfnamefont {D.}~\bibnamefont
  {Shao}}, \bibinfo {author} {\bibfnamefont {Q.}~\bibnamefont {Zhang}}, \ and\
  \bibinfo {author} {\bibfnamefont {P.}~\bibnamefont {Xu}},\ }\href@noop {}
  {\bibfield  {journal} {\bibinfo  {journal} {Polymers}\ }\textbf {\bibinfo
  {volume} {14}},\ \bibinfo {pages} {2134} (\bibinfo {year}
  {2022})}\BibitemShut {NoStop}%
\bibitem [{\citenamefont {Srivastava}\ \emph {et~al.}(2020)\citenamefont
  {Srivastava}, \citenamefont {Chattopadhyaya},\ and\ \citenamefont
  {Bandyopadhyay}}]{Srivastava2020}%
  \BibitemOpen
  \bibfield  {author} {\bibinfo {author} {\bibfnamefont {R.}~\bibnamefont
  {Srivastava}}, \bibinfo {author} {\bibfnamefont {M.}~\bibnamefont
  {Chattopadhyaya}}, \ and\ \bibinfo {author} {\bibfnamefont {P.}~\bibnamefont
  {Bandyopadhyay}},\ }\href {\doibase 10.1039/C9CP05578A} {\bibfield  {journal}
  {\bibinfo  {journal} {Phys. Chem. Chem. Phys.}\ }\textbf {\bibinfo {volume}
  {22}},\ \bibinfo {pages} {2142} (\bibinfo {year} {2020})}\BibitemShut
  {NoStop}%
\bibitem [{\citenamefont {Chen}\ \emph {et~al.}(2022)\citenamefont {Chen},
  \citenamefont {Nardi}, \citenamefont {Amadei},\ and\ \citenamefont
  {D’Abramo}}]{Chen2022}%
  \BibitemOpen
  \bibfield  {author} {\bibinfo {author} {\bibfnamefont {C.~G.}\ \bibnamefont
  {Chen}}, \bibinfo {author} {\bibfnamefont {A.~N.}\ \bibnamefont {Nardi}},
  \bibinfo {author} {\bibfnamefont {A.}~\bibnamefont {Amadei}}, \ and\ \bibinfo
  {author} {\bibfnamefont {M.}~\bibnamefont {D’Abramo}},\ }\href {\doibase
  10.3390/molecules27031077} {\bibfield  {journal} {\bibinfo  {journal}
  {Molecules}\ }\textbf {\bibinfo {volume} {27}} (\bibinfo {year} {2022}),\
  10.3390/molecules27031077}\BibitemShut {NoStop}%
\bibitem [{\citenamefont {Ullmann}\ and\ \citenamefont
  {Knapp}(1999)}]{Ullmann1999}%
  \BibitemOpen
  \bibfield  {author} {\bibinfo {author} {\bibfnamefont {G.}~\bibnamefont
  {Ullmann}}\ and\ \bibinfo {author} {\bibfnamefont {E.-W.}\ \bibnamefont
  {Knapp}},\ }\href@noop {} {\bibfield  {journal} {\bibinfo  {journal} {Eur.
  Biophys. J.}\ }\textbf {\bibinfo {volume} {28}},\ \bibinfo {pages} {533}
  (\bibinfo {year} {1999})}\BibitemShut {NoStop}%
\bibitem [{\citenamefont {Busch}\ \emph {et~al.}(2015)\citenamefont {Busch},
  \citenamefont {Houser}, \citenamefont {Hayden}, \citenamefont {Sherman},
  \citenamefont {Lafer},\ and\ \citenamefont {Stachowiak}}]{Busch2015}%
  \BibitemOpen
  \bibfield  {author} {\bibinfo {author} {\bibfnamefont {D.~J.}\ \bibnamefont
  {Busch}}, \bibinfo {author} {\bibfnamefont {J.~R.}\ \bibnamefont {Houser}},
  \bibinfo {author} {\bibfnamefont {C.~C.}\ \bibnamefont {Hayden}}, \bibinfo
  {author} {\bibfnamefont {M.~B.}\ \bibnamefont {Sherman}}, \bibinfo {author}
  {\bibfnamefont {E.~M.}\ \bibnamefont {Lafer}}, \ and\ \bibinfo {author}
  {\bibfnamefont {J.~C.}\ \bibnamefont {Stachowiak}},\ }\href@noop {}
  {\bibfield  {journal} {\bibinfo  {journal} {Nat. Commun.}\ }\textbf {\bibinfo
  {volume} {6}},\ \bibinfo {pages} {1} (\bibinfo {year} {2015})}\BibitemShut
  {NoStop}%
\bibitem [{\citenamefont {Zeno}\ \emph {et~al.}(2019)\citenamefont {Zeno},
  \citenamefont {Thatte}, \citenamefont {Wang}, \citenamefont {Snead},
  \citenamefont {Lafer},\ and\ \citenamefont {Stachowiak}}]{Zeno2019}%
  \BibitemOpen
  \bibfield  {author} {\bibinfo {author} {\bibfnamefont {W.~F.}\ \bibnamefont
  {Zeno}}, \bibinfo {author} {\bibfnamefont {A.~S.}\ \bibnamefont {Thatte}},
  \bibinfo {author} {\bibfnamefont {L.}~\bibnamefont {Wang}}, \bibinfo {author}
  {\bibfnamefont {W.~T.}\ \bibnamefont {Snead}}, \bibinfo {author}
  {\bibfnamefont {E.~M.}\ \bibnamefont {Lafer}}, \ and\ \bibinfo {author}
  {\bibfnamefont {J.~C.}\ \bibnamefont {Stachowiak}},\ }\href@noop {}
  {\bibfield  {journal} {\bibinfo  {journal} {J. Am. Chem. Soc.}\ }\textbf
  {\bibinfo {volume} {141}},\ \bibinfo {pages} {10361} (\bibinfo {year}
  {2019})}\BibitemShut {NoStop}%
\bibitem [{\citenamefont {Jin}\ \emph {et~al.}(2022)\citenamefont {Jin},
  \citenamefont {Cao},\ and\ \citenamefont {Baumgart}}]{Rui2022}%
  \BibitemOpen
  \bibfield  {author} {\bibinfo {author} {\bibfnamefont {R.}~\bibnamefont
  {Jin}}, \bibinfo {author} {\bibfnamefont {R.}~\bibnamefont {Cao}}, \ and\
  \bibinfo {author} {\bibfnamefont {T.}~\bibnamefont {Baumgart}},\ }\href@noop
  {} {\bibfield  {journal} {\bibinfo  {journal} {Sci. Rep.}\ }\textbf {\bibinfo
  {volume} {12}},\ \bibinfo {pages} {7676} (\bibinfo {year}
  {2022})}\BibitemShut {NoStop}%
\bibitem [{\citenamefont {Ferrenberg}\ and\ \citenamefont
  {Swendsen}(1989)}]{Ferrenberg1989}%
  \BibitemOpen
  \bibfield  {author} {\bibinfo {author} {\bibfnamefont {A.~M.}\ \bibnamefont
  {Ferrenberg}}\ and\ \bibinfo {author} {\bibfnamefont {R.~H.}\ \bibnamefont
  {Swendsen}},\ }\href@noop {} {\bibfield  {journal} {\bibinfo  {journal}
  {Phys. Rev. Lett.}\ }\textbf {\bibinfo {volume} {63}},\ \bibinfo {pages}
  {1195} (\bibinfo {year} {1989})}\BibitemShut {NoStop}%
\bibitem [{\citenamefont {Kumar}\ \emph {et~al.}(1992)\citenamefont {Kumar},
  \citenamefont {Rosenberg}, \citenamefont {Bouzida}, \citenamefont
  {Swendsen},\ and\ \citenamefont {Kollman}}]{Kumar1992}%
  \BibitemOpen
  \bibfield  {author} {\bibinfo {author} {\bibfnamefont {S.}~\bibnamefont
  {Kumar}}, \bibinfo {author} {\bibfnamefont {J.~M.}\ \bibnamefont
  {Rosenberg}}, \bibinfo {author} {\bibfnamefont {D.}~\bibnamefont {Bouzida}},
  \bibinfo {author} {\bibfnamefont {R.~H.}\ \bibnamefont {Swendsen}}, \ and\
  \bibinfo {author} {\bibfnamefont {P.~A.}\ \bibnamefont {Kollman}},\
  }\href@noop {} {\bibfield  {journal} {\bibinfo  {journal} {J. Comp. Chem.}\
  }\textbf {\bibinfo {volume} {13}},\ \bibinfo {pages} {1011} (\bibinfo {year}
  {1992})}\BibitemShut {NoStop}%
\bibitem [{\citenamefont {K\"stner}(2011)}]{Kastner2011}%
  \BibitemOpen
  \bibfield  {author} {\bibinfo {author} {\bibfnamefont {J.}~\bibnamefont
  {K\"stner}},\ }\href@noop {} {\bibfield  {journal} {\bibinfo  {journal}
  {Wiley Interdiscip. Rev. Comput. Mol. Sci.}\ }\textbf {\bibinfo {volume}
  {1}},\ \bibinfo {pages} {932} (\bibinfo {year} {2011})}\BibitemShut {NoStop}%
\bibitem [{\citenamefont {Gorgulla}\ \emph {et~al.}(2020)\citenamefont
  {Gorgulla}, \citenamefont {Boeszoermenyi}, \citenamefont {Wang},
  \citenamefont {Fischer}, \citenamefont {Coote}, \citenamefont
  {Padmanabha~Das}, \citenamefont {Malets}, \citenamefont {Dmytro},
  \citenamefont {Moroz}, \citenamefont {Scott}, \citenamefont {Fackeldey},
  \citenamefont {Hoffmann}, \citenamefont {Iavniuk}, \citenamefont {Wagner},\
  and\ \citenamefont {Arthanari}}]{Gorgulla2020}%
  \BibitemOpen
  \bibfield  {author} {\bibinfo {author} {\bibfnamefont {C.}~\bibnamefont
  {Gorgulla}}, \bibinfo {author} {\bibfnamefont {A.}~\bibnamefont
  {Boeszoermenyi}}, \bibinfo {author} {\bibfnamefont {Z.-F.}\ \bibnamefont
  {Wang}}, \bibinfo {author} {\bibfnamefont {P.}~\bibnamefont {Fischer}},
  \bibinfo {author} {\bibfnamefont {P.}~\bibnamefont {Coote}}, \bibinfo
  {author} {\bibfnamefont {K.}~\bibnamefont {Padmanabha~Das}}, \bibinfo
  {author} {\bibfnamefont {Y.}~\bibnamefont {Malets}}, \bibinfo {author}
  {\bibfnamefont {S.}~\bibnamefont {Dmytro}}, \bibinfo {author} {\bibfnamefont
  {Y.}~\bibnamefont {Moroz}}, \bibinfo {author} {\bibfnamefont
  {D.}~\bibnamefont {Scott}}, \bibinfo {author} {\bibfnamefont
  {K.}~\bibnamefont {Fackeldey}}, \bibinfo {author} {\bibfnamefont
  {M.}~\bibnamefont {Hoffmann}}, \bibinfo {author} {\bibfnamefont
  {I.}~\bibnamefont {Iavniuk}}, \bibinfo {author} {\bibfnamefont
  {G.}~\bibnamefont {Wagner}}, \ and\ \bibinfo {author} {\bibfnamefont
  {H.}~\bibnamefont {Arthanari}},\ }\href@noop {} {\bibfield  {journal}
  {\bibinfo  {journal} {Nature}\ }\textbf {\bibinfo {volume} {580}},\ \bibinfo
  {pages} {1} (\bibinfo {year} {2020})}\BibitemShut {NoStop}%
\bibitem [{\citenamefont {Nos{\'{e}}}(1984{\natexlab{a}})}]{Nose1984}%
  \BibitemOpen
  \bibfield  {author} {\bibinfo {author} {\bibfnamefont {S.}~\bibnamefont
  {Nos{\'{e}}}},\ }\href@noop {} {\bibfield  {journal} {\bibinfo  {journal}
  {Mol. Phys.}\ }\textbf {\bibinfo {volume} {52}},\ \bibinfo {pages} {255}
  (\bibinfo {year} {1984}{\natexlab{a}})}\BibitemShut {NoStop}%
\bibitem [{\citenamefont {Nos{\'{e}}}(1984{\natexlab{b}})}]{Nose1984b}%
  \BibitemOpen
  \bibfield  {author} {\bibinfo {author} {\bibfnamefont {S.}~\bibnamefont
  {Nos{\'{e}}}},\ }\href@noop {} {\bibfield  {journal} {\bibinfo  {journal} {J.
  Chem. Phys.}\ }\textbf {\bibinfo {volume} {81}},\ \bibinfo {pages} {511}
  (\bibinfo {year} {1984}{\natexlab{b}})}\BibitemShut {NoStop}%
\bibitem [{\citenamefont {Hoover}(1985)}]{Hoover1985}%
  \BibitemOpen
  \bibfield  {author} {\bibinfo {author} {\bibfnamefont {W.~G.}\ \bibnamefont
  {Hoover}},\ }\href@noop {} {\bibfield  {journal} {\bibinfo  {journal} {Phys.
  Rev. A}\ }\textbf {\bibinfo {volume} {31}},\ \bibinfo {pages} {1695}
  (\bibinfo {year} {1985})}\BibitemShut {NoStop}%
\bibitem [{\citenamefont {Shelley}\ and\ \citenamefont
  {Patey}(1995)}]{Shelley1995}%
  \BibitemOpen
  \bibfield  {author} {\bibinfo {author} {\bibfnamefont {J.~C.}\ \bibnamefont
  {Shelley}}\ and\ \bibinfo {author} {\bibfnamefont {G.~N.}\ \bibnamefont
  {Patey}},\ }\href@noop {} {\bibfield  {journal} {\bibinfo  {journal} {J.
  Chem. Phys.}\ }\textbf {\bibinfo {volume} {102}},\ \bibinfo {pages} {7656}
  (\bibinfo {year} {1995})}\BibitemShut {NoStop}%
\bibitem [{\citenamefont {Itoh}\ \emph {et~al.}(2011)\citenamefont {Itoh},
  \citenamefont {Damjanovi{\'c}},\ and\ \citenamefont {Brooks}}]{itoh2011}%
  \BibitemOpen
  \bibfield  {author} {\bibinfo {author} {\bibfnamefont {S.~G.}\ \bibnamefont
  {Itoh}}, \bibinfo {author} {\bibfnamefont {A.}~\bibnamefont
  {Damjanovi{\'c}}}, \ and\ \bibinfo {author} {\bibfnamefont {B.~R.}\
  \bibnamefont {Brooks}},\ }\href@noop {} {\bibfield  {journal} {\bibinfo
  {journal} {Proteins}\ }\textbf {\bibinfo {volume} {79}},\ \bibinfo {pages}
  {3420} (\bibinfo {year} {2011})}\BibitemShut {NoStop}%
\bibitem [{\citenamefont {Lee}\ \emph {et~al.}(2014)\citenamefont {Lee},
  \citenamefont {Miller}, \citenamefont {Damjanovic},\ and\ \citenamefont
  {Brooks}}]{lee2014}%
  \BibitemOpen
  \bibfield  {author} {\bibinfo {author} {\bibfnamefont {J.}~\bibnamefont
  {Lee}}, \bibinfo {author} {\bibfnamefont {B.~T.}\ \bibnamefont {Miller}},
  \bibinfo {author} {\bibfnamefont {A.}~\bibnamefont {Damjanovic}}, \ and\
  \bibinfo {author} {\bibfnamefont {B.~R.}\ \bibnamefont {Brooks}},\
  }\href@noop {} {\bibfield  {journal} {\bibinfo  {journal} {J. Chem. Theory
  Comp.}\ }\textbf {\bibinfo {volume} {10}},\ \bibinfo {pages} {2738} (\bibinfo
  {year} {2014})}\BibitemShut {NoStop}%
\bibitem [{\citenamefont {Akram}\ \emph {et~al.}(2014)\citenamefont {Akram},
  \citenamefont {Rehman},\ and\ \citenamefont {Hall}}]{Muhammad2014}%
  \BibitemOpen
  \bibfield  {author} {\bibinfo {author} {\bibfnamefont {M.}~\bibnamefont
  {Akram}}, \bibinfo {author} {\bibfnamefont {J.}~\bibnamefont {Rehman}}, \
  and\ \bibinfo {author} {\bibfnamefont {E.}~\bibnamefont {Hall}},\ }\href@noop
  {} {\bibfield  {journal} {\bibinfo  {journal} {Antioxid. Redox Signal.}\
  }\textbf {\bibinfo {volume} {7}} (\bibinfo {year} {2014})}\BibitemShut
  {NoStop}%
\bibitem [{\citenamefont {Sugita}\ \emph {et~al.}(2000)\citenamefont {Sugita},
  \citenamefont {Kitao},\ and\ \citenamefont {Okamoto}}]{sugita2000}%
  \BibitemOpen
  \bibfield  {author} {\bibinfo {author} {\bibfnamefont {Y.}~\bibnamefont
  {Sugita}}, \bibinfo {author} {\bibfnamefont {A.}~\bibnamefont {Kitao}}, \
  and\ \bibinfo {author} {\bibfnamefont {Y.}~\bibnamefont {Okamoto}},\
  }\href@noop {} {\bibfield  {journal} {\bibinfo  {journal} {J. Chem. Phys.}\
  }\textbf {\bibinfo {volume} {113}},\ \bibinfo {pages} {6042} (\bibinfo {year}
  {2000})}\BibitemShut {NoStop}%
\bibitem [{\citenamefont {Socci}\ \emph {et~al.}(1996)\citenamefont {Socci},
  \citenamefont {Onuchic},\ and\ \citenamefont {Wolynes}}]{Socci1996}%
  \BibitemOpen
  \bibfield  {author} {\bibinfo {author} {\bibfnamefont {N.~D.}\ \bibnamefont
  {Socci}}, \bibinfo {author} {\bibfnamefont {J.~N.}\ \bibnamefont {Onuchic}},
  \ and\ \bibinfo {author} {\bibfnamefont {P.~G.}\ \bibnamefont {Wolynes}},\
  }\href@noop {} {\bibfield  {journal} {\bibinfo  {journal} {J. Chem. Phys.}\
  }\textbf {\bibinfo {volume} {104}},\ \bibinfo {pages} {5860} (\bibinfo {year}
  {1996})}\BibitemShut {NoStop}%
\bibitem [{\citenamefont {Hummer}\ \emph {et~al.}(2000)\citenamefont {Hummer},
  \citenamefont {Garc\'{\i}a},\ and\ \citenamefont {Garde}}]{Hummer2000}%
  \BibitemOpen
  \bibfield  {author} {\bibinfo {author} {\bibfnamefont {G.}~\bibnamefont
  {Hummer}}, \bibinfo {author} {\bibfnamefont {A.~E.}\ \bibnamefont
  {Garc\'{\i}a}}, \ and\ \bibinfo {author} {\bibfnamefont {S.}~\bibnamefont
  {Garde}},\ }\href@noop {} {\bibfield  {journal} {\bibinfo  {journal} {Phys.
  Rev. Lett.}\ }\textbf {\bibinfo {volume} {85}},\ \bibinfo {pages} {2637}
  (\bibinfo {year} {2000})}\BibitemShut {NoStop}%
\end{thebibliography}%
%

\end{document}